%% file: main.tex
\documentclass[%
 aip,
pop,
showkeys,
 amsmath,amssymb,
reprint,%
onecolumn,
]{revtex4-2}

\usepackage{amsmath}
\usepackage{amsthm}
\usepackage{amssymb}
\usepackage{color}
\usepackage{xcolor}
\usepackage{cases}
\usepackage{enumitem}              
\usepackage{mathrsfs}
\usepackage{mathtools}
\usepackage{siunitx}
\usepackage[OMLmathsfit]{isomath}  
\usepackage{graphicx}
\usepackage{grffile}               
\usepackage[T1]{fontenc}
\usepackage{lmodern}

\usepackage{bm}                      
\usepackage{tikz}                    
\usetikzlibrary{arrows.meta, positioning, shapes.geometric, calc}
\usepackage{hyperref}                
\usepackage{cleveref}                

\newcommand{\vd}{\mathbf{\cdot}}                       
\providecommand{\grad}{\vec{\nabla}}                   
\newcommand{\half}[1][1]{\frac{#1}{2}}
\newcommand{\third}[1][1]{\frac{#1}{3}}

\providecommand{\tensor}{}\renewcommand{\tensor}[1]{\underline{\underline{#1}}}

\newcommand{\abs}[1]{\left\lvert #1 \right\rvert}
\newcommand{\upwind}[1]{\left\langle #1 \right\rangle}   
\newcommand{\density}{\ensuremath{\rho}}
\newcommand{\charge}{e}
\newcommand{\electric}{\epsilon_0}

\providecommand{\selectlanguage}[1]{}
\renewcommand{\selectlanguage}[1]{}

\makeatother

\begin{document}


%
%
\title{Beyond Radiation Hydrodynamics: A Coupled Kinetic Plasma and Radiation Modeling of NIF Polar Direct Drive Capsule}

\author{Hans R. Hammer}
\thanks{Corresponding author}
\email{hrhammer@lanl.gov}
\affiliation{Theoretical Division, Los Alamos National Laboratory, Los Alamos, NM, 87545, USA}

\author{William T. Taitano}
\affiliation{Theoretical Division, Los Alamos National Laboratory, Los Alamos, NM, 87545, USA}

\author{Steven E. Anderson}
\affiliation{Computational Physics Division, Los Alamos National Laboratory, Los Alamos, NM, 87545, USA}

\author{Brett D. Keenan}
\affiliation{Computational Physics Division, Los Alamos National Laboratory, Los Alamos, NM, 87545, USA}

\author{Luis Chac{\'o}n}
\affiliation{Theoretical Division, Los Alamos National Laboratory, Los Alamos, NM, 87545, USA}

\author{Andrei N. Simakov}
\affiliation{Theoretical Design Division, Los Alamos National Laboratory, Los Alamos, NM, 87545, USA}

%
%
%
%
%
%

\date{\today}

\begin{abstract}
Predictive modeling of inertial-confinement-fusion (ICF) implosions increasingly
requires resolving ion-kinetic effects beyond the reach of
radiation-hydrodynamics, without sacrificing radiation transport. We present, to
our knowledge, the first coupling of a kinetic-ion plasma model to thermal
radiation transport. The \texttt{iFP}
code---a conservative, spherically symmetric, 1D2V hybrid
kinetic-ion/fluid-electron Vlasov--Fokker--Planck solver---is extended with a
grey $P_1$ radiation model, coupled through a high-order/low-order (HOLO)
fixed-point solver and advanced on a shared adaptive moving mesh. Opacities are
mixed on the fly from the LANL \texttt{TOPS} library, exploiting the per-species
information intrinsic to a kinetic code. We verify the implementation against the
optically thick Marshak-wave and Mach-45 radiative-shock benchmarks, and apply it
to a NIF polar-direct-drive exploding-pusher capsule. Increasing physics fidelity
systematically improves agreement with the measured neutron yield and
burn width; the improvement is driven by kinetically enhanced fuel leakage and
ballistic preheating, and radiation transport proves essential for quantitative
agreement. This work establishes a proof-of-principle capability for higher-fidelity kinetic-plasma/radiation modeling of ICF.
\end{abstract}
\keywords{Vlasov--Fokker--Planck; Grey radiation transport; Polar Direct Drive; HOLO; \texttt{iFP}}

\maketitle

%
%
\input{sections/introduction}

%
%
\input{sections/model}

%
%
\input{sections/solver}

%
%
\input{sections/verification}

%
%

\input{sections/nif_pdd}

%
%
\input{sections/conclusions}

%
%
\section*{Acknowledgement}
\label{acknowledgement}

W.T.T was supported partially under the LANL office of experimental sciences between 2023-2026. This work was also sponsored by the Institutional Computing program at the Los Alamos National Laboratory and was performed under the auspices of the National Nuclear Security Administration of the U.S. Department of Energy at Los Alamos National Laboratory, managed by Triad National Security, LLC under contract 89233218CNA000001.

%
%
%
%
%
\appendix
\crefalias{section}{appendix}
\crefalias{subsection}{appendix}
\input{sections/appendix}

%
%
\bibliographystyle{apsrev4-2}
\bibliography{references}

%
%

\end{document}

%% file: sections/introduction.tex
\section{Introduction}
\label{sec:introduction}

Laser-driven inertial confinement fusion (ICF) uses laser (direct drive) or x-ray (indirect drive) radiation
to implode and ignite capsules containing thermonuclear fuel, typically a deuterium--tritium (DT) mixture.
In these high-energy-density (HED) experiments, the coupled dynamics of the plasma and the radiation field
govern the assembly, stagnation, and burn of the fuel~\cite{lindl_review_2014,
nationalresearchcouncil_assessment_2013}. Predictive modeling of these systems
must therefore self-consistently capture transport of mass, momentum, and
energy in the plasma together with its exchange with the radiation field across
density, temperature, and ion and photon mean-free-path ranges spanning many
orders of magnitude. The
fidelity of this modeling directly informs the design and interpretation of
experiments at facilities such as the National Ignition Facility (NIF) and OMEGA.

To date, the workhorse tools for ICF and HED design have been
radiation-hydrodynamics (rad-hydro) codes, which often treat the plasma as a single 
fluid and couple it to a thermal-radiation-transport (TRT)
model of varying sophistication~\cite{gittings_rage_2008, holst_crash:_2011,
haines_development_2022, mihalas_foundations_2013}. These codes have been
remarkably successful: they model the multi-material hydrodynamics, laser
energy deposition, and radiation transport that set the gross behavior of a
capsule implosion, and they underpin much of the current understanding of ICF
performance. The fluid closure employed by such codes, however, presumes that each particle species remains close to a local Maxwellian distribution, with short particle mean free paths
relative to the gradient scale lengths. This assumption breaks down in the
long-mean-free-path regime characteristic of low-fill-density, thin-shell
capsules, where ion-kinetic effects---species separation, non-local heat
transport, and the formation of non-Maxwellian tails---become
significant. Exploding-pusher targets fielded on OMEGA have provided
particularly clean evidence of such kinetic behavior~\cite{mannion_pre_2023_above_balabio}, and analyses of NIF
implosions have reported signatures of suprathermal ion distributions in
the burning plasma~\cite{hartouni_evidence_2023}. Together, these
observations motivate a high-fidelity kinetic description of the ions, for which the
Vlasov--Fokker--Planck (VFP) formulation, coupled to a self-consistent radiation treatment, forms the
appropriate first-principles model.

A number of fully kinetic and hybrid kinetic-ion/fluid-electron capabilities have been
developed to address this regime~\cite{taitano_cpc_2021,larroche2018omega}. 
What does not yet exist in the ICF modeling toolbox, however,
is a fully self-consistent kinetic-plasma capability that also includes
radiation transport and laser energy deposition---the ingredients required to
model a realistic laser-driven implosion from first principles. In this work, we
take a first step toward closing this gap by addressing radiation transport
(laser deposition is left to future work). We present the \texttt{iFP} kinetic
code---a conservative, spherically symmetric, 1D2V hybrid kinetic-ion/fluid-electron
VFP solver---extended here with a grey $P_1$ radiation model. The plasma and
radiation treatments are coupled through a high-order/low-order (HOLO) fixed-point
solver~\cite{chacon_multiscale_2017}, and material opacities are supplied
inline from the Los Alamos National Laboratory (LANL) \texttt{TOPS} opacity library, exploiting the per-species
information available in a kinetic code to mix opacities consistently. We regard
this coupled capability as a proof of principle: the grey $P_1$ closure is
deliberately simple, but it establishes the coupling machinery and quantifies,
for the first time to our knowledge in a kinetic-ion framework, the role of radiation in an ICF
implosion.

We demonstrate the capability by simulating a NIF polar-direct-drive (PDD) exploding-pusher
capsule~\cite{yeamans_high_2021}. PDD platforms have been proposed as high-yield
fusion neutron sources, and their comparatively low fill pressures and thin
shells push them further into the kinetic regime than higher-convergence
ignition capsules, amplifying the ion-kinetic effects discussed above.
Compared with OMEGA-scale exploding pushers, however, NIF PDD capsules reach
substantially larger convergence ratios and fuel densities, making radiation an
important energy-loss mechanism. Predictive modeling of these targets
therefore requires \emph{both} a kinetic treatment of the ions \emph{and} a
consistent treatment of the radiation field---precisely the combination this
work makes possible.

The remainder of this manuscript is organized as follows.
\Cref{sec:model} presents the physical model: the hybrid
kinetic-ion/fluid-electron VFP system in spherical geometry, its grey
$P_1$ radiation extension, and the adaptive moving-mesh formulation used to
track the imploding boundary and steep gradients.
\Cref{sec:solver} describes the semi-discrete time integration and the
nonlinearly implicit HOLO solution strategy, together with the opacity-mixing
model specialized to a kinetic plasma.
\Cref{sec:verification} verifies the implementation against the optically thick
Marshak-wave and Mach-45 radiative-shock benchmarks.
\Cref{sec:nif_pdd_modeling} applies the coupled capability to the NIF PDD
exploding-pusher capsule and analyzes the interplay of kinetic and radiative
effects on the implosion.
\Cref{sec:conclusions} summarizes our findings and outlines future directions.

%% file: sections/model.tex
\section{Model}
\label{sec:model}

We model the capsule as a spherically symmetric plasma composed of kinetic ions and fluid
electrons, coupled to a thermal radiation field. The ions are described by a
Vlasov--Fokker--Planck (VFP) kinetic equation in a 1D2V phase space---one
configuration-space dimension (the radius $r$) and two velocity-space
dimensions---while the electrons are treated as a massless fluid that enforces
quasi-neutrality. The radiation field is advanced with a grey
(frequency-integrated) $P_1$ model---coupled energy and flux moment equations---in
the non-relativistic limit. This hybrid kinetic-ion/fluid-electron
description follows the \texttt{iFP} formulation of
\citet{taitano_eulerian_2021, taitano_cpc_2021}; here we summarize the governing
equations and highlight the modifications required to couple the plasma to the
radiation field on an adaptive moving mesh. The moving-mesh transformation and
discretization of the radiation moments are detailed in \cref{app:moving_mesh}.

\subsection{Kinetic ions and fluid electrons}
\label{subsec:model:plasma}

We assume spherical symmetry in the
configuration space, so that the only spatial coordinate is the radius, $r$. Spherical symmetry further implies that $f$ is gyrotropic about $\hat{r}$, so the velocity is naturally decomposed into components parallel and perpendicular to $\hat{r}$,
\begin{equation}
    \label{eq:model:velocity_split}
    v_\| = \vec{v}\vd\hat{r},
    \qquad
    v_\perp = \sqrt{v^2 - v_\|^2},
\end{equation}
with $v_\|$ playing the role of the cylindrical axial coordinate, $v_\perp$ that
of the cylindrical radial coordinate, and $\phi_v$ the ignorable azimuthal angle.
Because the distribution function is independent of the two configuration-space
angles \emph{and} of $\phi_v$, the six-dimensional phase space collapses to
1D2V \emph{exactly}, rather than as an approximation~\cite{taitano_cpc_2021}.

Each ion species $\alpha$ is thus represented by a distribution function
$f_\alpha(r, v_\|, v_\perp, t)$ that satisfies the 1D2V electrostatic VFP
equation:
\begin{equation}
    \label{eq:model:ion:vfp}
    \partial_t (J_S f_\alpha) + \partial_r (J_S v_\| f_\alpha)
    + \frac{Z_\alpha e}{m_\alpha} E_\| \, \partial_{v_\|}(J_S f_\alpha)
    + \frac{1}{r}\,\partial_v \vd (J_S \tilde{\vec{a}} f_\alpha)
    = J_S \left(\sum_\beta C_{\alpha\beta} + C_{\alpha e}\right),
\end{equation}
where $J_S = r^2$ is the spherical Jacobian, $Z_\alpha e$ and $m_\alpha$ are the
charge and mass of species $\alpha$ with $e$ the elementary charge, $E_\|$ is the radial component of the
self-consistent electric field, $\tilde{\vec{a}} = [\,v_\perp^2,\, -v_\| v_\perp\,]^{\mathsf T}$
is the geometric (pseudo-)acceleration arising from the spherical/cylindrical
coordinates, and $\partial_v\vd\vec{A} = \partial_{v_\|} A_\| +
v_\perp^{-1}\partial_{v_\perp}(v_\perp A_\perp)$ is the reduced velocity-space
divergence. The collision term $C_{\alpha\beta}$ is the Fokker--Planck operator
for collisions of ion species $\alpha$ with ion species $\beta$; we adopt the
Rosenbluth--MacDonald--Judd form,
\begin{equation}
    \label{eq:model:ion:fp_operator}
    C_{\alpha\beta} = \Gamma_{\alpha\beta}\,\partial_v \vd
    \left( \tensor{D}_\beta \vd \grad_v f_\alpha - \vec{A}_\beta f_\alpha \right),
    \qquad
    \Gamma_{\alpha\beta} = \frac{Z_\alpha^2 Z_\beta^2 \charge^4 \Lambda_{\alpha\beta}}
    {8\pi\electric^2 m_\alpha^2},
\end{equation}
with $\Lambda_{\alpha\beta}$ the Coulomb logarithm and $\tensor{D}_\beta$,
$\vec{A}_\beta$ the velocity-space diffusion tensor and friction vector built
from the Rosenbluth potentials~\cite{rosenbluth_fokker-planck_1957}; their
numerical treatment is described in \citet{taitano_adaptive_2018}.

Collisions with the electrons are treated separately through $C_{\alpha e}$,
since the electrons are advanced as a fluid rather than carried as a kinetic
species [\cref{eq:model:electron:energy}]. Here,
\begin{equation}
    \label{eq:model:ion:fp_operator_electron}
    C_{\alpha e} = \nu_{\alpha e}\,\partial_v \vd
    \left( D_{\alpha e} \grad_v f_\alpha - \vec{A}_{\alpha e} f_\alpha \right),
\end{equation}
with the coefficients $D_{\alpha e}$, $\vec{A}_{\alpha e}$ and collision
frequency $\nu_{\alpha e}$ evaluated from a Maxwellian electron background
parameterized by $n_e$, $u_{e,\|}$, and $T_e$, rather than from a kinetic
distribution as in \cref{eq:model:ion:fp_operator}. This
closed-form Maxwellian-background evaluation is what makes it practical to
retain a massless fluid electron model while still resolving the ion
distribution kinetically; the corresponding moments of $C_{\alpha e}$ supply
the collisional friction $\mathcal{F}_{\|,\alpha e}$ and energy exchange $W_{e\alpha}$
appearing in the electron equations
[\cref{eq:model:electron:energy,eq:model:lo_electrical_field}]. A detailed treatment
of these terms is beyond the scope of this work; see
\citet{taitano_adaptive_2018}.

Quasi-neutrality and the current-free (ambipolar) condition give
\begin{align}
    \label{eq:model:lo_quasineutrality}
    \sum_\alpha Z_\alpha n_\alpha - n_e &= 0, \\
    \label{eq:model:lo_ambipolarity}
    \sum_\alpha Z_\alpha n_\alpha u_{\alpha,\|} - n_e u_{e,\|} &= 0,
\end{align}
which determine the electron number density $n_e$ and drift velocity $u_{e,\|}$
from the ion moments. In the spherically symmetric conservative form consistent
with \cref{eq:model:ion:vfp} (spherical Jacobian $J_S = r^2$), the electron
internal energy evolves as
\begin{equation}
    \label{eq:model:electron:energy}
    \half[3]\partial_t\!\left(J_S P_e\right)
    + \half[5]\partial_r\!\left(J_S u_{e,\|} P_e\right)
    + \partial_r\!\left(J_S Q_{e,\|}\right)
    - J_S\!\left(q_e n_e u_{e,\|} E_\| + \sum_\alpha W_{e\alpha} + S_{re}\right) = 0,
\end{equation}
where $P_e = n_e T_e$ is the electron pressure, $Q_{e,\|}$ the radial electron
heat flux, $W_{e\alpha}$ the collisional energy exchange with ion species
$\alpha$, and $S_{re}$ the energy exchanged with the radiation field
[\cref{eq:model:radiation:energy_deposition}]. Neglecting electron inertia
($m_e \to 0$), the electron momentum balance reduces to a generalized Ohm's law
for the electric field,
\begin{equation}
    \label{eq:model:lo_electrical_field}
    E_{\|} = - \frac{\partial_r P_e + \sum_\alpha \mathcal{F}_{\|,\alpha e} - S_{rp,\|}}{e n_e},
\end{equation}
with $\mathcal{F}_{\|,\alpha e}$ the radial ion--electron friction and $S_{rp,\|}$ the
radiation momentum deposition [\cref{eq:model:radiation:momentum_deposition}].
The transport coefficients $Q_{e,\|}$, $W_{e\alpha}$, and $\mathcal{F}_{\|,\alpha e}$
follow the multi-ion Braginskii closure of \citet{molvig_pop_2014}.

\subsection{Grey radiation $P_1$ equations}
\label{subsec:model:radiation}

The radiation field is advanced through its frequency-integrated energy $E_r$
and radial component of the radiation flux $F_\|$, which obey the grey $P_1$ moment equations (spherical Jacobian $J_S = r^2$;
the energy equation is in conservative form, the flux equation in
non-conservative form),
\begin{align}
    \label{eq:model:radiation:energy}
    \partial_t\!\left(J_S E_r\right) + \partial_r\!\left(J_S F_\|\right) + J_S S_{re} &= 0, \\
    \label{eq:model:radiation:momentum}
    \frac{1}{c^2}\partial_t\!\left(J_S F_\|\right) + \third J_S\,\partial_r E_r + J_S S_{rp, \|} &= 0,
\end{align}
where the radial component of the isotropic $P_1$ pressure gradient is $\third\partial_r E_r$ (the geometric terms of the spherical divergence
cancel for an isotropic radiation pressure). The energy and momentum exchange
with the plasma are
\begin{align}
    \label{eq:model:radiation:energy_deposition}
    S_{re} &= \sigma_a c E_0 - \sigma_e a c T_e^4 + \frac{\sigma_r}{c} u_{e, \|} F_{0, \|}, \\
    \label{eq:model:radiation:momentum_deposition}
    S_{rp, \|} &= \frac{\sigma_r}{c} F_{0, \|}
    + \frac{u_{e, \|}}{c} \left( \sigma_a E_0 - \sigma_e a T_e^4 \right),
\end{align}
where $\sigma_a$, $\sigma_e$, and $\sigma_r$ are the absorption, emission, and
transport (Rosseland) cross sections, $a$ is the radiation constant, and $c$ is the
speed of light. The comoving-frame energy and radial flux, $E_0$ and $F_{0,\|}$,
are related to their laboratory-frame counterparts by the material-motion
corrections~\cite{morel_discrete-ordinates_2006}
\begin{equation}
    \label{eq:model:radiation:comoving}
    E_0 = E_r - 2 \frac{u_{e, \|}}{c^2} F_{0, \|},
    \qquad
    F_{0, \|} = F_\| - \frac{4}{3} u_{e, \|} E_r .
\end{equation}

For the non-relativistic regime of interest ($u_{e,\|} \ll c$), we
retain the material-motion corrections only to leading order in $u_{e,\|}/c$, so that
the momentum deposition and comoving energy simplify to
\begin{equation}
    \label{eq:model:radiation:simplified}
    S_{rp, \|} \simeq \frac{\sigma_r}{c} F_{0, \|},
    \qquad
    E_0 \simeq E_r .
\end{equation}
However, the diffusion momentum deposition term
\begin{equation}
    \label{eq:model:radiation:diffusion_momentum}
    S_{rp, \|} = -\frac{1}{3}\partial_r E_r ,
\end{equation}
obtained by substituting the diffusion-limit flux
$F_{0,\|} = -\left(c/3\sigma_r\right)\partial_r E_r$ into
\cref{eq:model:radiation:simplified}, was found to yield the most robust coupling
to the plasma momentum balance through the electron equation
\cref{eq:model:lo_electrical_field}.

We solve the full $P_1$ system \cref{eq:model:radiation:energy,eq:model:radiation:momentum} via nonlinear elimination (\cref{app:mm:elimination}): $F_\|$ is evolved as an independent moment but is slaved to $E_r$ during the solve. The cross sections $\sigma_{a,e,r}$ are assembled on the fly from tabulated opacities using the mixing model of \cref{subsec:solver:opacity}.

\subsection{Adaptive moving mesh}
\label{subsec:model:moving_mesh}

An imploding capsule develops a converging boundary and steep, co-moving
gradients (shock fronts, ablation surfaces, the fuel--shell interface) that a
fixed Eulerian grid cannot resolve economically. We therefore solve the plasma
and radiation equations on an adaptive mesh defined by a time-dependent map
$r(\xi, t)$ from a fixed logical coordinate $\xi \in [0,1]$ to the physical
radius,
\begin{equation}
    \label{eq:model:mesh_map}
    r(\xi, t) = \int_0^\xi J_\xi(\xi', t)\, d\xi',
    \qquad J_\xi = \partial_\xi r .
\end{equation}
Relative to the plasma-only formulation of
\citet{taitano_cpc_2021}, the radiation energy and flux are transformed and
discretized on the same moving mesh (\cref{subsec:model:moving_mesh_equations}), so that the plasma and radiation subsystems
share a single grid throughout the implosion.

The mesh is positioned by a nonlinear Winslow-type moving-mesh PDE (MMPDE) driven
by a monitor function $\omega(\xi,t)$,
\begin{equation}
    \label{eq:model:mmpde}
    \partial_t r = \tau_r^{-1}\,\partial_\xi\!\left[\omega\,\partial_\xi r\right],
    \qquad
    r|_{\xi=0} = 0,\quad r|_{\xi=1} = R(t),
\end{equation}
where $\tau_r$ is a grid-relaxation time scale and $R(t)$ the (moving) outer
boundary. The monitor function concentrates mesh points where the solution
varies rapidly. Following \citet{taitano_cpc_2021} we build it from the inverse
gradient-length scales of the ion number density $n_\alpha[f_\alpha]$, thermal speed
$v_{th,\alpha}[f_\alpha]$, drift velocity $u_{\|,\alpha}[f_\alpha]$, and electron temperature $T_e$;
\emph{in this work we augment the monitor with the radiation temperature}, $T_r = (E_r/a)^{1/4}$, so that the
mesh also tracks the radiation front,
\begin{equation}
    \label{eq:model:monitor}
    \omega = \sqrt{\;\frac{1}{2}\sum_\alpha
    \left[\left(L_{n,\alpha}^{-1}\right)^2
    + \left(L_{v_{th},\alpha}^{-1}\right)^2
    + \left(L_{u_\|,\alpha}^{-1}\right)^2\right]
    + \left(L_{T_e}^{-1}\right)^2
    + \left(L_{T_r}^{-1}\right)^2\;},
\end{equation}
with each inverse length scale of the form $L_{q}^{-1} = |\delta_\xi \ln q| +
\delta_{\min}$ and $\delta_{\min}$ a small floor on the relative variation. The
$(\,\cdot\,)^{-1}$ length scales, the limiting and Winslow-smoothing of $\omega$,
and the nonlinear volume-preserving stabilization of the map are unchanged from
\citet{taitano_cpc_2021}, to which we refer the reader.

In addition to the configuration-space mesh, \texttt{iFP} carries a separate,
adaptive velocity-space grid for each ion species, since different species can
differ by orders of magnitude in instantaneous thermal spread and drift and can
evolve on comparably fast time scales. The physical parallel and perpendicular
velocities are related to logical (dimensionless) velocity-space coordinates
$\tilde{v}_\|, \tilde{v}_\perp$ through a per-species normalization speed
$v^*_\alpha(r,t)$ and drift offset $u^*_{\|,\alpha}(r,t)$,
\begin{equation}
    \label{eq:model:velocity_mesh_map}
    v_\| = v^*_\alpha\, \tilde{v}_\| + u^*_{\|,\alpha},
    \qquad
    v_\perp = v^*_\alpha\, \tilde{v}_\perp,
\end{equation}
so that the velocity grid expands or contracts with $v^*_\alpha$ and translates
with $u^*_{\|,\alpha}$ to track the local thermal speed and drift velocity of
species $\alpha$---the velocity-space analogue of the configuration-space map
\cref{eq:model:mesh_map}. As with $r(\xi,t)$, the metrics $v^*_\alpha$ and
$u^*_{\|,\alpha}$ are not prescribed but are themselves advanced in time,
following the adaptation strategy of \citet{taitano_cpc_2021}.
Because \cref{eq:model:velocity_mesh_map} makes the velocity coordinates
both space- and time-dependent, transforming the ion VFP equation to these
logical velocity variables introduces an inertial term (a pseudo-acceleration),
whose two contributions are, respectively, the expansion/contraction of the
velocity grid as $v^*_\alpha$ evolves and its rigid translation as
$u^*_{\|,\alpha}$ evolves. Although this term is essential for a practical
Eulerian simulation of the implosion, we omit its detailed treatment here and
refer the reader to \citet{taitano_eulerian_2021}.

\subsection{Moving-mesh form of the governing equations}
\label{subsec:model:moving_mesh_equations}

Transforming the plasma and radiation equations to the logical coordinate $\xi$
introduces the mesh Jacobian $J_\xi = \partial_\xi r$ and the grid velocity
$\dot{r} \equiv \partial_t r|_\xi$, and replaces material velocities in the advective fluxes by their
values relative to the moving grid. Writing the transport in
arbitrary-Lagrangian--Eulerian (ALE) conservative form on the composite
phase-space volume element (spherical Jacobian $J_S = r^2$, mesh Jacobian
$J_\xi$), the ion VFP equation \cref{eq:model:ion:vfp} becomes
\begin{equation}
    \label{eq:model:mm:vfp}
    \partial_t\!\left(J_\xi J_S f_\alpha\right)
    + \partial_\xi\!\left[J_S\left(v_\| - \dot{r}\right) f_\alpha\right]
    + \frac{Z_\alpha e}{m_\alpha} E_\|\,\partial_{v_\|}\!\left(J_\xi J_S f_\alpha\right)
    + \frac{J_\xi}{r}\,\partial_v\!\vd\!\left(J_S \tilde{\vec{a}} f_\alpha\right)
    = J_\xi J_S \left(\sum_\beta C_{\alpha\beta} + C_{\alpha e}\right).
\end{equation}
The grid motion thus enters only through the relative advection velocity
$v_\| - \dot{r}$; the field and geometric acceleration terms of
\cref{eq:model:ion:vfp} are simply rescaled by $J_\xi$.
The electron internal-energy equation \cref{eq:model:electron:energy} takes the
ALE form
\begin{equation}
    \label{eq:model:mm:electron_energy}
    \half[3]\partial_t\!\left(J_\xi J_S\, P_e\right)
    + \partial_\xi\!\left[J_S\!\left(\left(u_{e,\|} - \dot{r}\right)\half[3] P_e
    + u_{e,\|} P_e + Q_{e,\|}\right)\right]
    - J_\xi J_S\!\left(q_e n_e u_{e,\|} E_\| + \sum_\alpha W_{e\alpha} + S_{re}\right) = 0,
\end{equation}
and the grey radiation $P_1$ energy and flux equations
\cref{eq:model:radiation:energy,eq:model:radiation:momentum} take the ALE form
\begin{align}
    \label{eq:model:mm:radiation_energy}
    \partial_t\!\left(J_\xi J_S\, E_r\right)
    + \partial_\xi\!\left[J_S\!\left(F_\| - \dot{r}\, E_r\right)\right]
    + J_\xi J_S\, S_{re} &= 0, \\
    \label{eq:model:mm:radiation_flux}
    \frac{1}{c^2}\partial_t\!\left(J_\xi J_S\, F_\|\right)
    - \frac{1}{c^2}\partial_\xi\!\left(J_S\, \dot{r}\, F_\|\right)
    + \third J_S\,\partial_\xi E_r
    + J_\xi J_S\, S_{rp, \|} &= 0,
\end{align}
so that both radiation moments are advanced on the same moving mesh as the
plasma. As in \cref{eq:model:radiation:momentum}, the isotropic radiation
pressure enters the flux equation as the gradient $\third J_S\,\partial_\xi E_r$
rather than a conservative divergence.
\Cref{eq:model:mm:vfp,eq:model:mm:electron_energy,eq:model:mm:radiation_energy,eq:model:mm:radiation_flux}
are the final system advanced by the solver of \cref{sec:solver}; their detailed
derivation, including the geometric source terms and the treatment of the outer
moving boundary, is given in \cref{app:moving_mesh}.

%% file: sections/solver.tex
\section{Numerical Method}
\label{sec:solver}

The coupled kinetic-ion/fluid-electron/radiation system derived in
\cref{sec:model} is advanced monolithically with a fully nonlinearly implicit time
integrator. This is a deliberate choice: the collisional, radiative, and
mesh-motion terms span disparate time scales---the ion--electron collision
time and the radiation absorption--emission time are both orders of magnitude
shorter than the hydrodynamic time near stagnation---so an operator-split or
explicit treatment would be either unstable or prohibitively restrictive on
$\Delta t$. Solving all subsystems together also lets the discretization
preserve the conservation and equilibrium symmetries that the continuum
formulation was built to respect. Below we state the semi-discrete BDF2 system
(\cref{subsec:solver:bdf2}), describe the
opacity-mixing model that supplies the radiation cross sections
(\cref{subsec:solver:opacity}), and outline the high-order/low-order (HOLO) strategy
used to solve it efficiently (\cref{subsec:solver:holo}). A detailed discussion
of the HOLO treatment of the Rosenbluth--Fokker--Planck operator is beyond
the scope of this work and will be reported separately.

\subsection{Semi-discrete BDF2 formulation}
\label{subsec:solver:bdf2}

The moving-mesh equations
\cref{eq:model:mm:vfp,eq:model:mm:electron_energy,eq:model:mm:radiation_energy,eq:model:mm:radiation_flux}
are integrated in time with a second-order backward-differentiation
formula (BDF2)~\cite{byrne_bdf2_1975}. Writing the discrete time level as a
superscript $l$ and the step size as $\Delta t^{l}$, the BDF2 approximation of a
temporal derivative $\partial_t(\,\cdot\,)$ is the three-level stencil
\begin{equation}
    \label{eq:solver:bdf2_stencil}
    \partial_t \psi \;\longrightarrow\;
    \frac{1}{\Delta t^{l}} \sum_{l'=-1}^{1} a^{\,l+l'}\, \psi^{\,l+l'},
\end{equation}
where the coefficients $a^{\,l+l'}$ are the (variable-step) BDF2 weights at time
level $l$~\cite{byrne_bdf2_1975} and reduce to first-order backward Euler for
the start-up step. Hereafter, $a$ with a time-level index denotes these BDF2
weights, distinct from the radiation constant $a$ of \cref{eq:model:radiation:energy_deposition}. All spatial-flux, field, collision, and radiation--matter
coupling terms are evaluated implicitly at the new time level $l+1$, so that
each governing equation takes the residual form
$\sum_{l'} a^{\,l+l'}\psi^{\,l+l'} + \Delta t^{l}\,\mathcal{R}[\,\cdot\,]^{l+1}=0$,
with $\mathcal{R}$ collecting the non-temporal terms of the corresponding
moving-mesh equation. Applied to the ion VFP equation \cref{eq:model:mm:vfp},
this yields
\begin{equation}
    \label{eq:solver:discrete_vfp}
    \sum_{l'=-1}^{1} a^{\,l+l'}\!\left(J_\xi J_S f_\alpha\right)^{l+l'}
    + \Delta t^{l}\!\left\{
    \partial_\xi\!\left[J_S(v_\| - \dot{r}) f_\alpha\right]
    + \frac{Z_\alpha e}{m_\alpha} E_\|\,\partial_{v_\|}\!\left(J_\xi J_S f_\alpha\right)
    + \frac{J_\xi}{r}\,\partial_v\!\vd\!\left(J_S \tilde{\vec{a}} f_\alpha\right)
    - J_\xi J_S \left(\sum_\beta C_{\alpha\beta} + C_{\alpha e}\right)
    \right\}^{l+1} = 0,
\end{equation}
the electron internal-energy equation \cref{eq:model:mm:electron_energy} to
\begin{equation}
    \label{eq:solver:discrete_electron_energy}
    \half[3]\sum_{l'=-1}^{1} a^{\,l+l'}\!\left(J_\xi J_S\, P_e\right)^{l+l'}
    + \Delta t^{l}\!\left\{
    \partial_\xi\!\left[J_S\!\left((u_{e,\|}-\dot{r})\half[3]P_e + u_{e,\|}P_e + Q_{e,\|}\right)\right]
    - J_\xi J_S\!\left(q_e n_e u_{e,\|} E_\| + \sum_\alpha W_{e\alpha} + S_{re}\right)
    \right\}^{l+1} = 0,
\end{equation}
and the grey radiation $P_1$ energy equation
\cref{eq:model:mm:radiation_energy} to
\begin{equation}
    \label{eq:solver:discrete_radiation_energy}
    \sum_{l'=-1}^{1} a^{\,l+l'}\!\left(J_\xi J_S\, E_r\right)^{l+l'}
    + \Delta t^{l}\!\left\{
    \partial_\xi\!\left[J_S(F_\| - \dot{r}\,E_r)\right] + J_\xi J_S\, S_{re}
    \right\}^{l+1} = 0.
\end{equation}
The companion flux equation \cref{eq:model:mm:radiation_flux} discretizes to
\begin{equation}
    \label{eq:solver:discrete_radiation_flux}
    \frac{1}{c^2}\sum_{l'=-1}^{1} a^{\,l+l'}\!\left(J_\xi J_S\, F_\|\right)^{l+l'}
    + \Delta t^{l}\!\left\{
    -\frac{1}{c^2}\partial_\xi\!\left(J_S\,\dot{r}\,F_\|\right)
    + \third J_S\,\partial_\xi E_r + J_\xi J_S\, S_{rp,\|}
    \right\}^{l+1} = 0.
\end{equation}
The radiation flux $F_\|$ is not carried as an independent unknown in the
implicit solve. Once the
BDF2 stencil is applied, the new-time flux enters
\cref{eq:solver:discrete_radiation_flux} only algebraically---its sole nonlocal
coupling is the weak mesh-motion advection $\propto\dot{r}$---so the equation
can be solved for $F_\|^{\,l+1}$ in terms of the radiation energy, the electron
velocity, and the flux history, and the result substituted into
\cref{eq:solver:discrete_radiation_energy}. This nonlinearly eliminates $F_\|$
and \emph{enslaves} it to the radiation energy, leaving $E_r$ as the only
independent radiation unknown carried in the implicit solve. The elimination
shrinks the radiation block and improves its conditioning; in Newton form it is
a Schur complement on the $(E_r, F_\|)$ block, with the flux increment
$\delta F_\|$ given by a linearized closure and back-substituted into the energy
residual. The resulting closed-form flux, its linearization, and the tridiagonal
solve induced by the mesh-motion advection are derived in
\cref{app:mm:elimination} [\cref{eq:app:p1_flux,eq:app:linearized_flux}]. The
face-based discretization of the radiation moments is detailed in
\cref{app:mm:discretization}.

The cell-face positions are advanced concurrently by discretizing the MMPDE
\cref{eq:model:mmpde} with the same BDF2 stencil
\cref{eq:solver:bdf2_stencil},
\begin{equation}
    \label{eq:solver:discrete_mmpde}
    \sum_{l'=-1}^{1} a^{\,l+l'}\, r^{\,l+l'}
    = \frac{\Delta t^{l}}{\tau_r}\,
    \partial_\xi\!\left[\omega^{\,l+1}\,\partial_\xi r^{\,l+1}\right],
    \qquad
    r^{\,l+1}\big|_{\xi=0} = 0,\quad
    r^{\,l+1}\big|_{\xi=1} = R^{\,l+1},
\end{equation}
so that the new-time face positions $r^{\,l+1}(\xi)$---and hence the grid
velocity $\dot{r} = (\Delta t^{l})^{-1}\sum_{l'} a^{\,l+l'} r^{\,l+l'}$ recovered
from the same stencil---are converged together with the plasma and radiation
unknowns. Within each HOLO iteration the mesh is Picard-lagged, i.e., evaluated
from the previous HOLO iterate (\cref{subsec:solver:holo}). For the spatial discretization and monitor-function evaluation, see
\citet{taitano_eulerian_2021}.

\subsection{Opacity mixing}
\label{subsec:solver:opacity}

The radiation cross sections $\sigma_a$, $\sigma_e$, and $\sigma_r$ that close
the deposition terms \cref{eq:model:radiation:energy_deposition,eq:model:radiation:momentum_deposition}
are material properties with strongly nonlinear dependence on the local plasma
state. A distinguishing advantage of a kinetic-ion code over a single-fluid
rad-hydro code is that \texttt{iFP} carries the per-species number densities and
a common electron temperature explicitly, which allows the opacities to be
mixed on the fly from single-species data. We restrict attention to opacities
in local thermodynamic equilibrium (LTE); \texttt{iFP} can also treat non-LTE
opacities, but that capability is beyond the present scope.

We use a mass-weighted mixing model for all three cross-section types; since
they are treated identically, we suppress the type subscript in this
subsection. A cross section $\sigma$ is related to an opacity $\kappa$ by
\begin{equation}
    \label{eq:solver:xs_opacity}
    \sigma = \density\,\kappa,
\end{equation}
with $\density$ the mass density. Each ion species $\alpha$ carries its own
opacity $\kappa_\alpha(T_e, \density_{\mathrm{eff},\alpha})$, a function of the
electron temperature and the species' effective mass density
\begin{equation}
    \label{eq:solver:xs_effective_density}
    \density_{\mathrm{eff},\alpha} = \frac{n_e}{\bar{Z}_\alpha}\, m_\alpha,
\end{equation}
where $m_\alpha$ is the species mass and $\bar{Z}_\alpha$ is the mean ionization
(average charge state) of species $\alpha$---the average number of
electrons stripped per ion---which sets the electron density through the
quasineutrality relation $n_e = \sum_\alpha \bar{Z}_\alpha n_\alpha$
[\cref{eq:model:lo_quasineutrality}] and plays the role of the charge number
$Z_\alpha$ appearing in the ion VFP equation \cref{eq:model:ion:vfp}. The
mass-weighted opacity is
\begin{equation}
    \label{eq:solver:xs_weighted_opacity}
    \kappa = \sum_\alpha \eta_\alpha \kappa_\alpha,
\end{equation}
where $\eta_\alpha$ is the mass fraction of species $\alpha$. Combining
\cref{eq:solver:xs_opacity,eq:solver:xs_weighted_opacity} gives the mixed cross
section as the species opacities weighted by their partial mass densities,
\begin{equation}
    \label{eq:solver:xs_mixed}
    \sigma = \sum_\alpha \density_\alpha \kappa_\alpha,
    \qquad
    \density_\alpha = m_\alpha n_\alpha.
\end{equation}

\texttt{iFP} supports analytic opacities of the power-law form
\begin{equation}
    \label{eq:solver:xs_analytic}
    \kappa_\alpha = \kappa_{0,\alpha}\, \density_{\mathrm{eff},\alpha}^{\,b_\alpha}\, T_e^{\,g_\alpha},
\end{equation}
with model parameters $\kappa_{0,\alpha}$, $b_\alpha$, $g_\alpha$, as well as tabulated
\texttt{TOPS} data. The tables are read with \texttt{Gandolf}~\cite{judd_gandolf_2001} and stored
in memory per opacity type and species; a value at a given
$(T_e,\density_{\mathrm{eff},\alpha})$ is obtained by bisection to the
bracketing table entries followed by bilinear interpolation in log--log space.

The tabulated data also supply the species ionization state $\bar{Z}_\alpha$,
which itself depends on temperature and density. Through
\cref{eq:model:lo_quasineutrality,eq:solver:xs_effective_density} this
introduces a circular dependence between $\bar{Z}_\alpha$, $n_e$, and
$\density_{\mathrm{eff},\alpha}$; the electron temperature follows from
the evolved electron pressure through $T_e = P_e/n_e$. We
therefore solve the coupled system
\begin{align}
    \label{eq:solver:charge_system}
    \density_{\mathrm{eff},\alpha} = \frac{n_e}{\bar{Z}_\alpha}\,m_\alpha,
    \qquad
    T_e = \frac{P_e}{n_e},
    \qquad
    \bar{Z}_\alpha = \mathcal{Z}\!\left(T_e, \density_{\mathrm{eff},\alpha}\right),
    \qquad
    n_e = \sum_\alpha n_\alpha \bar{Z}_\alpha,
\end{align}
where $\mathcal{Z}$ is the ionization supplied by the opacity model or table
lookup. Because the tabulated $\bar{Z}_\alpha$ has no closed form, we solve
\cref{eq:solver:charge_system} cell-by-cell with the Picard iteration of
Algorithm~1, iterating to the convergence criterion
\begin{equation}
    \label{eq:solver:charge_tolerance}
    \frac{\abs{n_e^{k} - n_e^{k-1}}}{n_e^{k}} \le \tau_{Z},
\end{equation}
with $k$ the iteration index and $\tau_Z$ the tolerance; tabulated
$\bar{Z}_\alpha$ values are interpolated in log--log space as for the opacities.
Once the charge state has converged, the cross sections follow directly from
\cref{eq:solver:xs_mixed}. The opacities and charge states are updated at the
start of each time step using the previous step's state and then held fixed
through the step, consistent with the frozen-coefficient treatment invoked in
\cref{subsec:solver:holo}.

\par\medskip
\noindent\rule{\linewidth}{0.8pt}\par\vspace{2pt}
\noindent\textbf{Algorithm~1}\quad Average-charge solve in a single cell\par\vspace{2pt}
\noindent\rule{\linewidth}{0.4pt}\par\nobreak\vspace{2pt}
\begin{enumerate}[leftmargin=2em,itemsep=1pt,topsep=1pt,label=\arabic*.]
    \item Initialize from the previous time level $l$:
    $n_e^{0} \gets n_e^{\,l}$, $\;T_e^{0} \gets P_e^{\,l}/n_e^{\,l}$, and
    $\bar{Z}_\alpha^{0} \gets \bar{Z}_\alpha^{\,l}$ for all species $\alpha$.
    \item For $k = 1, 2, \dots, k_{\max}$:
    \begin{enumerate}[leftmargin=2em,itemsep=1pt,topsep=1pt,label=(\alph*)]
        \item set $n_e^{k} \gets 0$;
        \item for each species $\alpha$: update the effective density
        $\density_{\mathrm{eff},\alpha} \gets m_\alpha\, n_e^{k-1}/\bar{Z}_\alpha^{k-1}$,
        obtain $\bar{Z}_\alpha^{k}$ from the opacity model/table lookup
        $\mathcal{Z}(T_e^{k-1}, \density_{\mathrm{eff},\alpha})$, and accumulate
        $n_e^{k} \gets n_e^{k} + n_\alpha \bar{Z}_\alpha^{k}$;
        \item update the temperature $T_e^{k} \gets P_e/n_e^{k}$;
        \item if $\abs{n_e^{k} - n_e^{k-1}}/n_e^{k} \le \tau_{Z}$, stop.
    \end{enumerate}
\end{enumerate}
\vspace{-2pt}\noindent\rule{\linewidth}{0.8pt}\par\medskip

\subsection{HOLO solution strategy}
\label{subsec:solver:holo}

We briefly outline the high-order/low-order (HOLO) solver strategy employed to
solve the fully discretized, coupled nonlinear algebraic system of equations
arising from our model. Readers not
interested in these details may skip ahead to the NIF PDD simulation and
analysis in \cref{sec:nif_pdd_modeling}.
Equations \cref{eq:solver:discrete_vfp,eq:solver:discrete_electron_energy,eq:solver:discrete_radiation_energy}
(the last carrying the enslaved radiation flux
\cref{eq:solver:discrete_radiation_flux}),
together with the discrete quasineutrality, ambipolarity, and Ohm's-law
closures \cref{eq:model:lo_quasineutrality,eq:model:lo_ambipolarity,eq:model:lo_electrical_field}
and the MMPDE \eqref{eq:solver:discrete_mmpde}, constitute a single large nonlinear algebraic system per time
step. The dimensionality is dominated by the 1D2V ion distribution
functions, while the stiffest couplings---ion--electron collisions and
radiation absorption/emission---act on the low-dimensional fluid and radiation
moments. We exploit this structure with a high-order/low-order (HOLO)
algorithm~\cite{chacon_multiscale_2017}.

The high-order (HO) system is the kinetic ion equation
\cref{eq:solver:discrete_vfp}. The low-order (LO) system is the set of
velocity-space moments of \cref{eq:solver:discrete_vfp}---continuity,
momentum, and energy for each ion species---closed with the fluid-electron and
radiation equations and augmented by consistency terms evaluated from the HO
distribution, which force the LO moments to agree discretely with the moments
of the kinetic solution~\cite{chacon_multiscale_2017}. Because the electric
field, the collisional exchange, and the radiation coupling all enter through
this compact, moment-based LO system, the expensive HO kinetic solve is
decoupled from the stiff nonlinear coupling: the LO system is iterated to
convergence to accelerate the fixed point, and the HO distribution is updated
against the converged low-order fields. Here we treat the HOLO accelerator
as the underlying nonlinear solver and focus on the radiation coupling it
carries. Within a time step the opacities and charge states are held fixed at
values assembled from the previous step, as described next.

We now make the HO/LO decomposition precise. Collect the new-time ($l+1$)
unknowns into a high-order kinetic state $f = \{f_\alpha\}_{\alpha=1}^{N_s}$ over
the $N_s$ ion species and a low-order state
$U_{LO} = \{\,\mathcal{U}_{LO},\, P_e,\, E_r \}$,
where $\mathcal{U}_{LO} = \left\{\mathcal{U}_{LO,\alpha}\right\}_{\alpha=1}^{N_s}$ are
the LO ion mass, momentum, and energy densities. Their HO counterparts are the
velocity moments of the distribution, $\mathcal{U}_{HO,\alpha} = \langle \bm{\psi}_{\alpha}\, f_\alpha\rangle_v$ 
with the weight vector
$\bm{\psi}_{\alpha} = (1,\, m_\alpha v_\|,\, \half m_\alpha v^2)^{\mathsf T}$ and
$\langle\,\cdot\,\rangle_v$ denotes velocity-space integration.
The two subsystems are written in time semi-discrete form abstractly as
\begin{align}
    \label{eq:solver:ho_abstract}
    \text{(HO)}\qquad & a^{l+1} J^{l+1}_\xi J^{l+1}_S f^{(p+1)}_\alpha + \sum_{l'=-1}^{0} a^{\,l+l'}\!\left(J_\xi J_S\, f_{\alpha}\right)^{l+l'} + \Delta t^l \mathcal{H}_\alpha\!\left(f^{(p+1)}_\alpha;\, U^{(p)}_{LO}\right) = 0,
    \qquad \alpha = 1,\dots,N_s, \\
    \label{eq:solver:lo_abstract}
    \text{(LO)}\qquad & a^{l+1} J^{l+1}_\xi J^{l+1}_S U^{(p+1)}_{LO} + \sum_{l'=-1}^{0} a^{\,l+l'}\!\left(J_\xi J_S\, U_{LO}\right)^{l+l'} + \Delta t^l \mathcal{L}\!\left(U^{(p+1)}_{LO};\, \bm{\gamma}^{(p+1)}\right) = 0 .
\end{align}
The non-temporal operators in the HO equation, $\mathcal{H}_\alpha$, are from the discrete VFP equation
\cref{eq:solver:discrete_vfp}; the argument $U_{LO}$ signifies that the electric
field $E_\|$, and the stiff Rosenbluth--Fokker--Planck collision coefficients are \emph{parameterized} by the
LO iterate $U_{LO}^{(p)}$ and held fixed while
\cref{eq:solver:ho_abstract} is solved for $f^{(p+1)}_\alpha$. 
The non-temporal operators of the LO system, $\mathcal{L}$, collect the ion velocity-moment equations, the electron
internal-energy equation \cref{eq:solver:discrete_electron_energy}, and the radiation
energy equation \cref{eq:solver:discrete_radiation_energy} (carrying the enslaved
flux). The two systems are coupled through the consistency term $\bm{\gamma} = \left\{ \gamma_{\alpha} \right\}_{\alpha=1}^{N_s}$.
Taking the $\bm{\psi}$-moment of the non-temporal operators of the HO equation and subtracting the low-order
models (without the consistency terms) defines
\begin{equation}
    \label{eq:solver:consistency}
    \gamma^{(p)}_\alpha \equiv \gamma[f^{(p)}_{\alpha}] \;=\;
    \left\langle \bm{\psi}_{\alpha}\,,\, \mathcal{H}_\alpha (f^{(p)}_{\alpha};U^{(p)}_{HO})\right\rangle_v
    \;-\; \mathcal{L}_\alpha\!\left(U^{(p)}_{HO} ; 0 \right),
\end{equation}
that is, the discrepancy between the \emph{exact} kinetic moment and its
low-order representation. By construction $\gamma_\alpha$ vanishes when the
LO moments reproduce the HO moments discretely, so \cref{eq:solver:consistency}
acts as a closure: it injects the exact kinetic stress, heat flux, and
collisional-moment information---which the truncated fluid-moment hierarchy
cannot represent in closed form---into the LO system as a source term. Further, 
at the fully discrete level, the consistency term also accounts for the spatial 
discretization mismatch between the two systems. Crucially, $\gamma_\alpha$ is \emph{Picard-lagged}: it is evaluated from the
current HO iterate and frozen while the LO system is solved.

With this closure, the LO system \cref{eq:solver:lo_abstract} is itself a
\emph{fully nonlinearly coupled} algebraic system---the ion moments, electron
energy, and radiation energy are advanced simultaneously by a
Jacobian-free Newton--Krylov (JFNK)~\cite{knoll_jcp_2004_JFNK} solver to a relative tolerance $\epsilon_{LO}$---
but in drastically reduced dimension: per cell it carries only
$\mathcal{O}(N_s)$ fluid moments together with the electron and radiation unknowns,
rather than the full 1D2V distribution. It therefore captures the stiff,
nonlinear collisional and radiative coupling that would otherwise dictate the
time-step restriction, at a small fraction of the cost of the kinetic solve. The
expensive HO solve, in turn, sees only the transport structure of
\cref{eq:solver:discrete_vfp} with its stiff collision operators evaluated from
the (converged) LO moments and radiation fields and is therefore linear in $f_\alpha$ apart from the nonlinear (limited) flux
reconstruction, which requires inner nonlinear iterations to a tolerance
$\epsilon_{HO}$.
The MMPDE is deliberately excluded from the LO system: its coupling to the LO
unknowns is non-stiff, so it is Picard-lagged and advanced alongside the HO
solve.

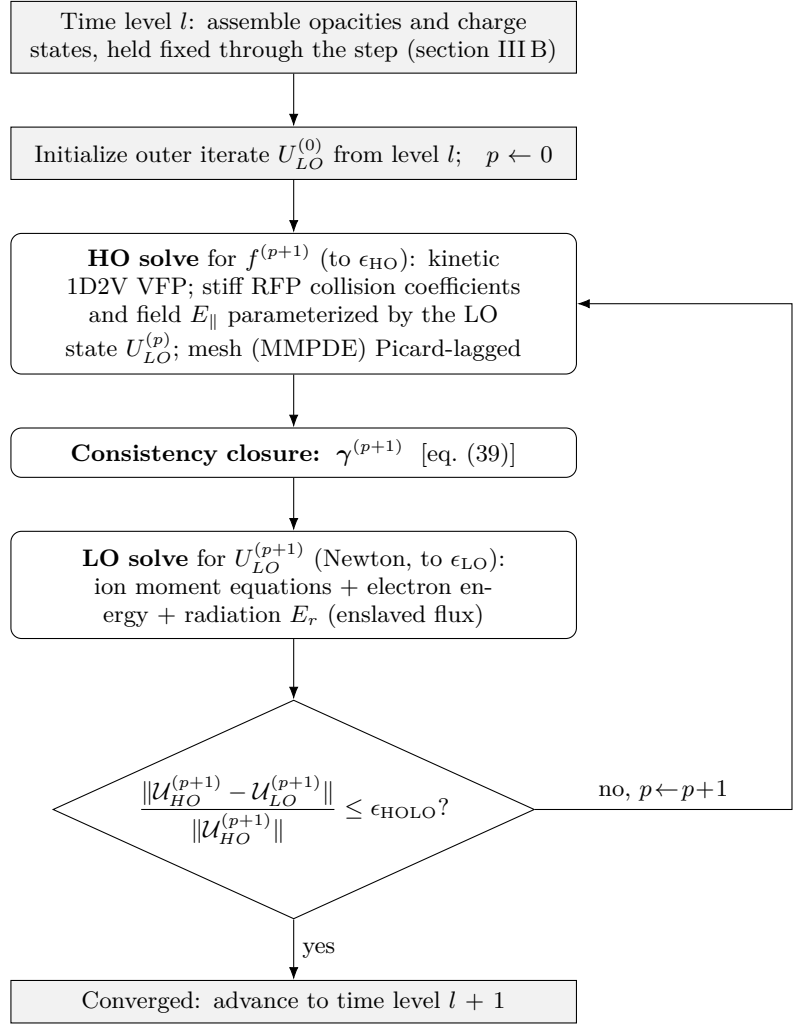
\begin{figure}[h!]
    \centering
    \begin{tikzpicture}[
        node distance = 7mm and 10mm,
        box/.style  = {rectangle, rounded corners, draw, align=center,
                       inner sep=4pt, text width=72mm, font=\small},
        term/.style = {rectangle, draw, align=center, inner sep=4pt,
                       text width=72mm, font=\small, fill=black!5},
        dec/.style  = {diamond, aspect=2.2, draw, align=center, inner sep=1pt,
                       font=\small},
        >=Latex
    ]
        \node[term] (start) {Time level $l$: assemble opacities and charge states, held fixed through the step (\cref{subsec:solver:opacity})};
        \node[term, below=of start] (init) {Initialize outer iterate $U_{LO}^{(0)}$ from level $l$;\quad $p \gets 0$};
        \node[box, below=of init] (ho) {\textbf{HO solve} for $f^{(p+1)}$ (to $\epsilon_{\mathrm{HO}}$): kinetic 1D2V VFP; stiff RFP collision coefficients and field $E_\|$ parameterized by the LO state $U_{LO}^{(p)}$; mesh (MMPDE) Picard-lagged};
        \node[box, below=of ho] (gamma) {\textbf{Consistency closure:}\ \ $\bm{\gamma}^{(p+1)}$ \ [\cref{eq:solver:consistency}]};
        \node[box, below=of gamma] (lo) {\textbf{LO solve} for $U^{(p+1)}_{LO}$ (Newton, to $\epsilon_{\mathrm{LO}}$): ion moment equations $+$ electron energy $+$ radiation $E_r$ (enslaved flux)};
        \node[dec, below=8mm of lo] (conv) {$\dfrac{\|\mathcal{U}_{HO}^{(p+1)}-\mathcal{U}_{LO}^{(p+1)}\|}{\|\mathcal{U}_{HO}^{(p+1)}\|}\le\epsilon_{\mathrm{HOLO}}$?};
        \node[term, below=8mm of conv] (done) {Converged: advance to time level $l+1$};

        \draw[->] (start) -- (init);
        \draw[->] (init) -- (ho);
        \draw[->] (ho) -- (gamma);
        \draw[->] (gamma) -- (lo);
        \draw[->] (lo) -- (conv);
        \draw[->] (conv) -- node[right, font=\small] {yes} (done);
        \draw[->] (conv.east) -- node[above, font=\small, align=center] {no,\ $p\!\gets\!p\!+\!1$} ++(34mm,0) |- (ho.east);
    \end{tikzpicture}
    \caption{One \texttt{iFP} time step as a HOLO fixed-point iteration. The
    high-order kinetic solve (converged to $\epsilon_{\mathrm{HO}}$) and the
    reduced low-order moment/radiation Newton solve (to $\epsilon_{\mathrm{LO}}$)
    are coupled through the Picard-lagged consistency closure $\bm{\gamma}$
    [\cref{eq:solver:consistency}]. The opacities and charge states are frozen at
    their start-of-step values, and the mesh (MMPDE) is Picard-lagged with the HO
    solve rather than carried in the LO system. The outer iteration repeats until
    the HO and LO moments agree to $\epsilon_{\mathrm{HOLO}}$.}
    \label{fig:holo_flow}
\end{figure}

The complete step is the fixed-point iteration sketched in
\cref{fig:holo_flow}. Given the low-order state $U_{LO}^{(p)}$ at outer iterate
$p$, we
\begin{enumerate}[leftmargin=2em,itemsep=1pt,topsep=2pt,label=(\roman*)]
    \item solve the HO kinetic system for each
    species (to a tolerance $\epsilon_{HO}$), with the stiff collision
    coefficients and the electric field $E_\|$ parameterized by the low-order
    state $U_{LO}^{(p)}$, and advance the mesh (MMPDE) Picard-lagged alongside;
    \item form the consistency terms
    $\bm{\gamma}^{(p+1)}$ from the updated HO moments
    $\mathcal{U}_{HO}^{(p+1)}$ [\cref{eq:solver:consistency}];
    \item solve the coupled nonlinear LO system by Newton
    iteration (to a tolerance $\epsilon_{LO}$) for the updated ion moments,
    electron energy, and radiation energy;
\end{enumerate}
iterating on $p$ until the HO and LO moments are mutually consistent,
$\|\mathcal{U}_{HO}^{(p+1)} - \mathcal{U}_{LO}^{(p+1)}\| / \|\mathcal{U}_{HO}^{(p+1)}\|
\le \epsilon_{\mathrm{HOLO}}$, at which point the step is converged.

To accelerate convergence of the outer fixed-point iteration in $p$, the
HOLO cycle sketched in \cref{fig:holo_flow} is wrapped in Anderson
acceleration~\cite{anderson_1965,walker_ni_2011}. The fixed-point variable
being accelerated is the low-order state $U_{LO}^{(p)}$---i.e., the
acceleration acts on the outer LO iterate rather than on the HO distribution
function---using a history depth of five.

%% file: sections/verification.tex
\section{Verification}
\label{sec:verification}

We verify the coupled implementation against two standard radiative-transfer
benchmarks that exercise complementary regimes of the model. The optically
thick Marshak wave (\cref{subsec:verification:marshak}) probes the diffusive,
strongly collisional limit in which the radiation and electron temperatures
equilibrate, stressing the radiation--electron energy coupling and the
adaptive mesh. The Mach-45 radiative shock (\cref{subsec:verification:shock})
probes the opposite regime of a strong, steady radiative shock with a radiative
precursor, testing the momentum and energy coupling with the material-motion corrections
\cref{eq:model:radiation:comoving}. Both problems are single-material, single-ion
tests; because \texttt{iFP} always carries the full ion kinetics, each case
requires adapting the kinetic model to reproduce the intended reference, as
described below.

\subsection{Marshak wave}
\label{subsec:verification:marshak}

We first consider the grey, optically thick Marshak-wave problem of
\citet{park_multigroup_2019}, run in planar geometry, in which a radiation
front propagates into cold material with a temperature-dependent opacity
$\sigma(T_e) = \density\,\kappa_0 / T_e^{3}$, used for all three cross
sections ($\sigma_a = \sigma_e = \sigma_r = \sigma$), where
$\kappa_0 = \SI{1e12}{\electronvolt\cubed\centi\meter\squared\per\gram}$. The calculation uses a grid of
$(N_x, N_\|, N_\perp) = (128,128,64)$, a domain $x \in [0, \SI{0.25}{\centi\meter}]$, $(\tilde{v}_\|,\tilde{v}_\perp) \in [-6,6]\times[0,6]$, and a time
window $t \in [0, \SI{20}{\nano\second}]$, with an initial time step $\Delta t_0 = \SI{1}{\femto\second}$ ramped to $\Delta t_{\max} = \SI{1}{\pico\second}$. The boundary source has a temperature of $T_{B} = \SI{1}{\kilo\electronvolt}$.  

\begin{figure}[h!]
    \centering
    \includegraphics[width=0.5\linewidth]{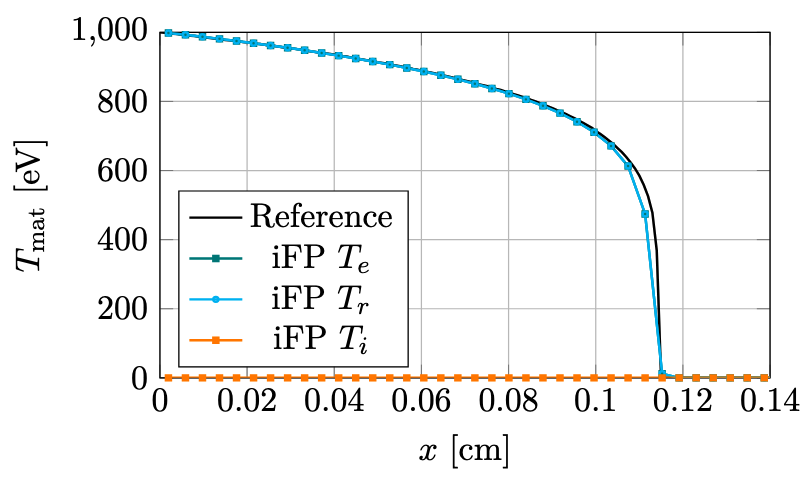}
    \caption{Grey optically thick Marshak-wave problem compared with the
    reference solution of \citet{park_multigroup_2019}. The \texttt{iFP}
    electron ($T_e$) and radiation ($T_r$) temperatures coincide and track the
    reference; the ion temperature $T_i$ is held artificially constant.}
    \label{fig:verification:marshak}
\end{figure}

The strongly collisional, diffusive regime is demanding for the kinetic ion
solver. Because \texttt{iFP} cannot disable the ion physics---in particular the ion
material motion---we emulate the static, electron-only Marshak problem by
rendering the ions dynamically inert: the ion mass is increased by a factor of 1000 and
the electron--ion Coulomb logarithm is reduced, suppressing energy flow into the
ions and freezing their motion on the time scale of the radiation front.
\Cref{fig:verification:marshak} compares the resulting material temperature
against the reference solution.
The electron and radiation temperatures are coincident,
as expected in the optically thick limit, so a single reference curve is shown.
The agreement is good; the small discrepancy in the wave-front shape is
likely due to the $P_1$ closure relative to the diffusion reference and to
the coarser \texttt{iFP} mesh.

\subsection{Mach-45 radiative shock}
\label{subsec:verification:shock}

The second benchmark is the Mach-45 radiative shock of
\citet{lowrie_radiative_2008}, run in planar geometry, which tests the momentum and energy coupling
between the plasma and radiation together with the material-motion corrections.
The solution is a steady traveling wave: the shock front translates while preserving its
profile. As this is originally a radiation-hydrodynamics test, the reference
provides only material and radiation moments, so we again constrain the kinetic
solver to reproduce a single-fluid state. We enforce strong ion--electron
coupling by artificially raising the electron--ion Coulomb logarithm to
$\Lambda_{\alpha e} = \num{1e6}$, which equilibrates the ion and electron
temperatures; the electron density and drift velocity are already tied to the
ions through the quasineutrality and ambipolarity closures
\cref{eq:model:lo_quasineutrality,eq:model:lo_ambipolarity}. The electron heat
conduction is suppressed with a heat-flux limiter of \num{1e-8}. The simulation
uses a mesh of $(N_x, N_\|, N_\perp) = (128,128,64)$ and is advanced to \SI{70}{\nano\second} with an initial time step $\Delta t_0 = \SI{0.1}{\pico\second}$ ramped to $\Delta t_{\max} = \SI{10}{\pico\second}$. The simulation is initialized from a given shock profile ($\rho, u_\|, T_e, E_r$) by interpolation onto the \texttt{iFP} cell centers, which covers a \SI{5}{\meter} section around the shock. Zero-gradient boundary conditions are applied. The material heat capacity is $c_v = \SI{1.4467e+12}{\text{erg}\per\gram\per\electronvolt}$. The absorption cross section has the form $\sigma_a = \SI{4.4940e+08}{\centi\meter^5\per\gram^2 \electronvolt^{\frac{7}{2}}} \times \frac{\rho^2}{T_e^{\frac{7}{2}}}$ and the scattering cross section is $\sigma_s = \SI{0.4006}{\centi\meter^2\per\gram} \times \rho$. The hydrodynamic reference state is mapped to plasma variables assuming $Z = 1$ and an ideal-gas equation of state.
The solver tolerances are $(\epsilon_{HOLO},\epsilon_{HO},\epsilon_{LO}) = (10^{-3}, 2\times10^{-2},10^{-4})$.

\begin{figure}[t]
    \centering
    \includegraphics[width=\linewidth]{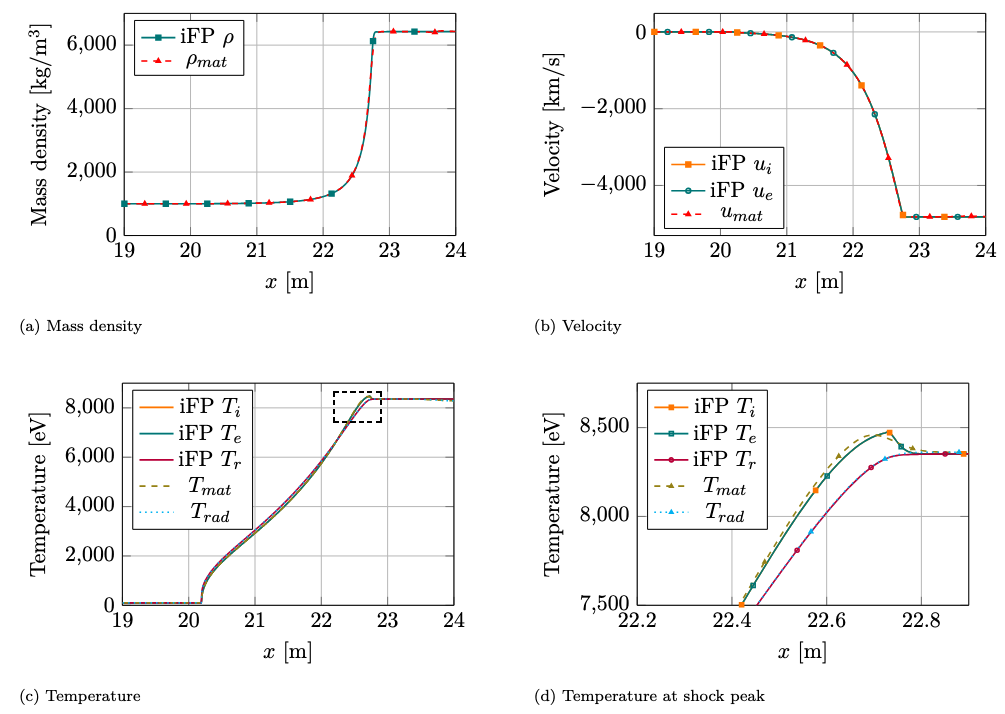}
    \caption{Mach-45 radiative-shock problem at \SI{70}{\nano\second} compared
    with the reference solution of \citet{lowrie_radiative_2008}. Top row:
    mass density (left) and ion/electron velocities (right). Bottom row:
    ion, electron, and radiation temperatures (left) and a detail of the
    temperature at the shock peak (right), corresponding to the dashed box in
    the bottom-left panel. Solid curves are \texttt{iFP};
    dashed/dotted curves are the reference.}
    \label{fig:verification:shock}
\end{figure}

\Cref{fig:verification:shock} compares the \texttt{iFP} mass density, velocity,
and temperature profiles against the reference. The agreement is good across all
moments; the only visible deviation is a minor discrepancy in the material
temperature near the trailing edge of the shock front, shown in the detail
panel (bottom right) of \cref{fig:verification:shock}.

\begin{figure}[h!]
    \centering 
    \includegraphics[width=0.92\linewidth]{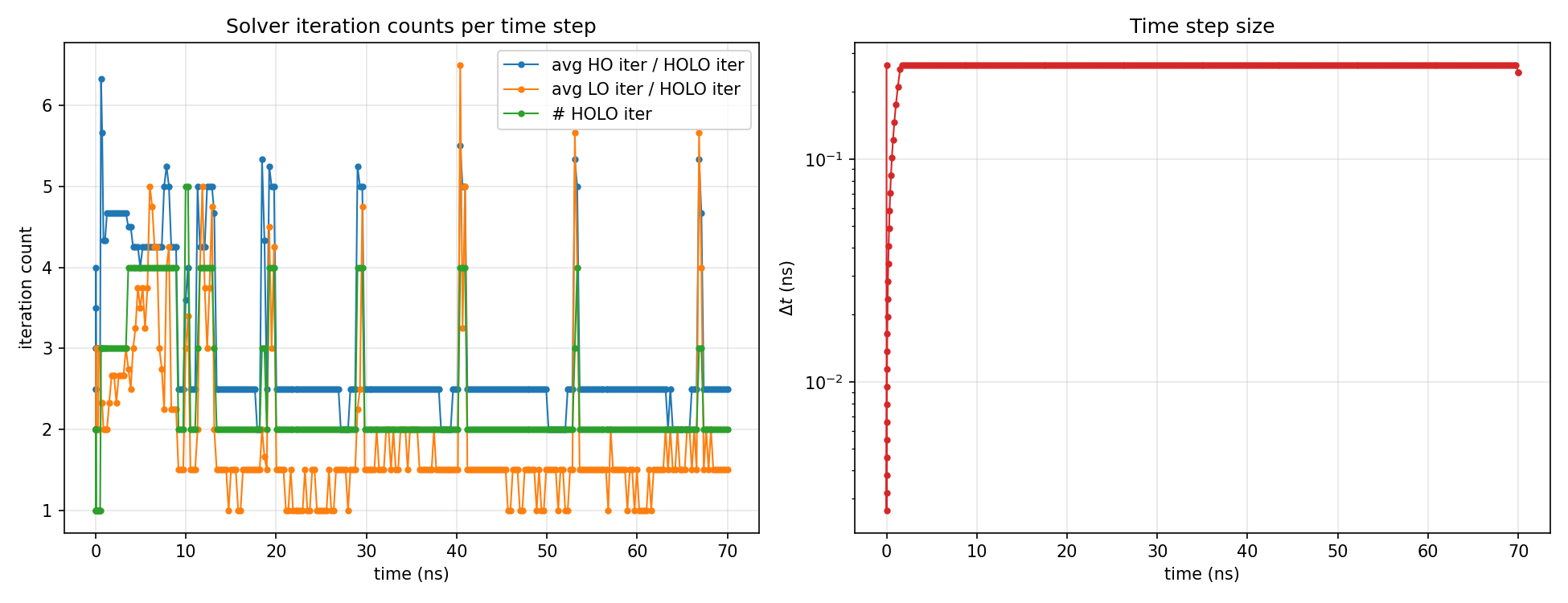}
    \caption{Solver statistics for the Mach-45 problem. Left: average HO and LO iterations per time step (blue and orange, respectively) and the number of HOLO iterations per time step (green). Right: time-step size $\Delta t$ as a function of time.}
    \label{fig:m45_verification:solver_stats}
\end{figure}

Solver statistics for this problem are shown in \cref{fig:m45_verification:solver_stats}.
The number of solver iterations remains bounded even for $\Delta t \gg \tau_{\mathrm{col}}$, underscoring the efficiency of the proposed method, where $\tau_{\mathrm{col}} \sim \SI{1e-5}{\nano\second}$ is the approximate collisional relaxation time scale of the cold, unshocked, and highly collisional plasma regions at $T = \SI{100}{\electronvolt}$ and $n \sim \SI{3e29}{\per\cubic\meter}$.

%% file: sections/nif_pdd.tex
\section{NIF Polar-Direct-Drive Modeling}
\label{sec:nif_pdd_modeling}
We demonstrate the integrated capability of the algorithm by modeling a realistic ICF experiment: the NIF PDD exploding-pusher capsule. The ``Cutie'' capsule was the largest of the three designs considered in \citet{yeamans_high_2021}, and it is a demanding test of kinetic-plasma/radiation coupling.

\begin{figure}[t]
    \centering
    \includegraphics[width=0.6\linewidth]{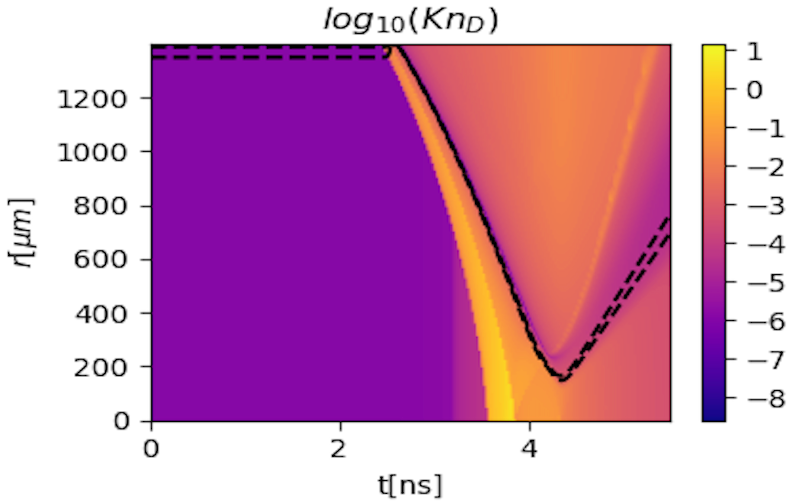}
    \caption{Deuterium Knudsen number radius--time (RT) streak plot obtained from the \texttt{xRAGE} simulation of the NIF PDD capsule. The black dashed lines denote the 90\% and 98\% fuel-mass-fraction contours, marking the fuel--shell interface.}
    \label{fig:knudsen_from_xrage}
\end{figure}

At early times, the high-temperature and low-density conditions in the fuel support strong ion-kinetic effects, which can significantly alter the implosion dynamics relative to hydrodynamic predictions. This is evidenced by the ion Knudsen number obtained from the \texttt{xRAGE} calculation of this capsule (to be discussed shortly); see \cref{fig:knudsen_from_xrage}.
As shown, the Knudsen number is of order unity or greater for the fuel species throughout the implosion stage, emphasizing the significance of ion-kinetic effects. Simultaneously, the dense and relatively cold shell exhibits large opacity, necessitating a coupled radiation and kinetic plasma treatment. Toward the end of the implosion, the fuel density rises rapidly and radiative losses become a leading term in the energy balance, so that a plasma-only description becomes inadequate. 

Modeling the Cutie capsule is particularly challenging, as it requires long simulation runtimes and demands solvers capable of robustly integrating over the stiff radiation and plasma-collision time scales encountered in the high-$Z$ ablator region. This makes it an ideal candidate for testing the algorithm's ability to model realistic multiscale implosion problems.

The Cutie capsule parameters, along with experimental and rad-hydro simulation performance metrics, are summarized in \cref{tab:cutie_summary}.
\begin{table}[t]
    \centering
    \begin{tabular}{llcr}
        \hline
        Capsule & Outer radius &($\mu$m) & 1978 \\
                 & Thickness &($\mu$m) & 22.7 \\
                 & Capsule mass &(mg) & 1.09 \\
        \hline
        Laser & Laser energy &(kJ) & 1256 \\
               & Ramp length &(ns) & 1.70 \\
               & Pulse length &(ns) & 3.90 \\
               & Peak power &(TW) & 425 \\
        \hline
        Gas fill & Fill pressure &(torr) & 5871 \\
                 & Fuel D atomic fraction &-& 0.6446 \\
                 & Initial fuel density &(mg/cm$^3$) & 1.52 \\
        \hline
        Experimental results & Yield &(DT neutrons) & $1.19\times10^{16}$ \\
                             & Ion temperature &(keV) & 10.22 \\
                             & Bang time &(ns) & 4.36 \\
                             & Burn width &(ps) & 421 \\
        \hline
        Postshot 1D \texttt{HYDRA} & Yield &(DT neutrons) & $1.09\times10^{17}$ \\
                                   & Ion temperature &(keV) & 13.46 \\
                                   & Bang time &(ns) & 3.83 \\
                                   & Burn width &(ps) & 196 \\
        \hline
    \end{tabular}
    \caption{Summary of the Cutie design parameters and results from \citet{yeamans_high_2021}.}
    \label{tab:cutie_summary}
\end{table}
As shown, the post-shot 1D \texttt{HYDRA} simulation overpredicts the yield by approximately an order of magnitude and the ion temperature by about 30\%. Here, the ion temperature is inferred from the full width at half-maximum (FWHM) of the DT neutron energy spectrum, measured experimentally via neutron time-of-flight (NTOF) detectors and computed analogously from the simulated neutron spectrum~\cite{yeamans_high_2021}. This discrepancy is commonly observed in exploding-pusher capsules even when asymmetries are small and fall-line analysis indicates negligible hydrodynamic mix. As discussed in detail below, the primary mechanisms of yield degradation in exploding pushers (relative to rad-hydro models) are fuel leakage from the core into the pusher via Knudsen-layer depletion, kinetically enhanced preheating of the hot spot, and the resulting reduction in convergence.

\citet{yeamans_high_2021} report that implementing ion diffusion--mix models and multigroup radiation diffusion in the \texttt{ARES} code improves yield agreement with experiments to within $\sim$30\%. However, we caution that while the diffusion--mix model qualitatively captures interatomic mixing physics, the mean free path in these capsules is too large for quantitative validity. Moreover, the \texttt{ARES} model relies on simplifying assumptions, such as a Lorentz-model thermal diffusivity coefficient, whose validity is questionable independent of kinetic effects~\cite{mackay_modeling_2020}.

We note that several physics packages are still absent from the \texttt{iFP} code for fully self-consistent, end-to-end NIF capsule simulations. These include (1) a laser package, (2) non-ideal ion EoS models, and (3) multigroup radiation transport. These effects are especially important in the ablated shell region and during early times when the shell is cold and dense: non-ideality and Fermi degeneracy modify the ion EoS and electron heat conduction, while multigroup radiation transport is needed because the frequency-dependent opacity in the shell varies significantly and the local radiation spectrum is expected to depart strongly from a Planckian. We emphasize that the grey $P_1$ model used here is not intended as a substitute for multigroup transport; its purpose is instead to capture the leading-order radiation--plasma energy exchange and momentum coupling. We flag departure from a Planckian spectrum as a known limitation of the grey treatment and return to it, along with a path toward multigroup extension, in \cref{sec:conclusions}.

For these reasons, we employ a two-step strategy. In the first step, we model the capsule using the \texttt{xRAGE} rad-hydro code, which incorporates these capabilities~\cite{gittings_rage_2008,haines_development_2022}. The resulting \texttt{xRAGE} simulation output is then used to initialize and drive the \texttt{iFP} simulation at a point in time when non-ideal effects become less significant (to be discussed shortly). Furthermore, the \texttt{iFP} simulation domain is truncated in the shell region to avoid explicitly modeling the laser energy deposition physics. Instead, the \texttt{xRAGE} simulation provides a time-dependent boundary condition at this cutoff location, driving the remainder of the simulation. 

This two-step coupling strategy is consistent with prior modeling approaches for OMEGA experiments and consistent with prior \texttt{iFP}~\cite{taitano_pop_2018, mannion_pre_2023_above_balabio, gatujohnson2024pre, reichelt2025pre, keenan_pop_2020_revolver_sim} and \texttt{FPion}~\cite{larroche2018omega, larroche2003kinetic_omega, larroche2003kinetic_rygg} modeling of ICF experiments. Because of this staged process, we carefully detail the setup of both simulations, paying particular attention to the \texttt{xRAGE} setup, which dictates the zeroth-order hydrodynamic evolution that drives the \texttt{iFP} simulation.

\subsection{\texttt{xRAGE} simulation setup}
\label{subsec:xrage_simulation}
\begin{figure}[t]
    \centering
    \includegraphics[scale=0.2]{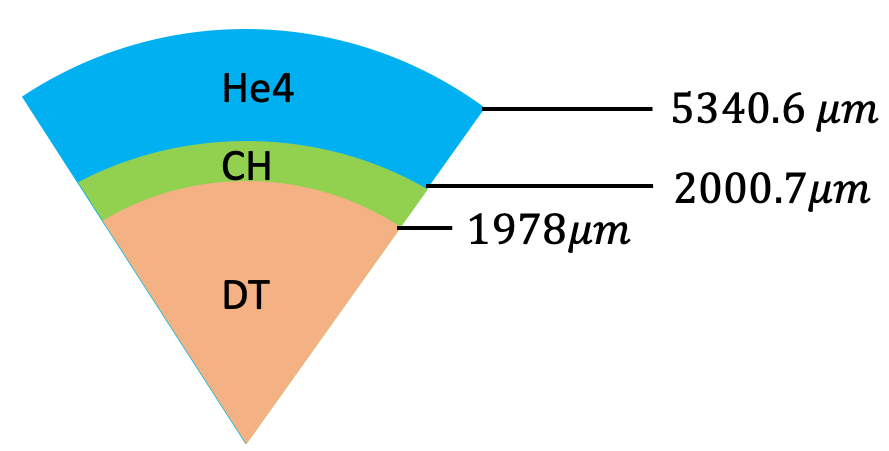}
    \caption{Schematic of the \texttt{xRAGE} zoning for the Cutie capsule. The capsule consists of the CH shell and DT fuel; the $^{4}$He background gas accommodates the ablated shell.}
    \label{fig:xrage_cutie_pie_chart}
\end{figure}

The 1D \texttt{xRAGE} simulation for the Cutie capsule models the DT fuel, CH shell, and $^{4}$He background gas as three distinct zones; see \cref{fig:xrage_cutie_pie_chart}.
\noindent The \texttt{xRAGE} simulation employs the following physics models:
\begin{itemize}
    \item the flux-limited Simakov--Molvig model for multi-ion-species electron,~\cite{simakov_electron_2014} run without a Fermi-degeneracy correction and using \texttt{xRAGE}'s default Coulomb logarithm model; 
    \item a generalized Lee-More model for the ions; 
    \item laser ray tracing with non-uniform energy deposition, using the \texttt{Mazinisin} laser package from LLE~\cite{marozas_pop_2006_mazinisin, marozas_prl_2018_mazinisin, marozas_pop_2018_mazinisin} in the 1D-in-2D option with 500 rays;
    \item a grey radiation diffusion model---the closest single-group radiation treatment available in \texttt{xRAGE} to the grey $P_1$ model used in \texttt{iFP}, though the two closures differ (a caveat we return to in \cref{sec:conclusions});
    \item an equation of state (EoS) from the \texttt{SESAME} library, with a material-concentration-based splitting between the electron and ion contributions;
    \item a 21-moment Grad-type multi-species ion diffusion model~\cite{vold_multispecies_2019}; and
    \item local thermodynamic equilibrium (LTE) opacities from the \texttt{TOPS} library.
\end{itemize}

\begin{figure}[t]
    \centering
    \includegraphics[scale=0.45]{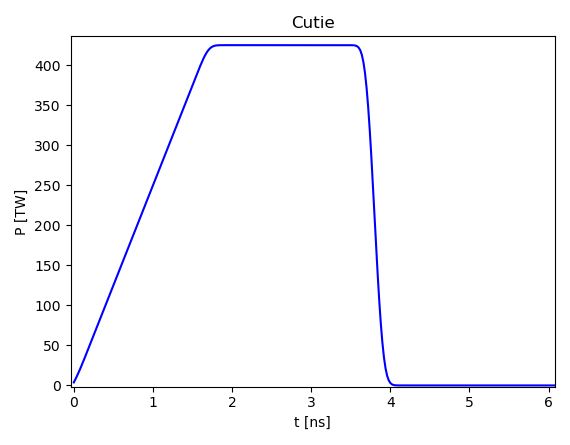}
    \caption{Laser power profile used in the \texttt{xRAGE} simulation of the Cutie capsule, based on the ramp-up, ramp-down, and peak power parameters reported by \citet{yeamans_high_2021}.}
    \label{fig:laser_profile}
\end{figure}

The laser power profile was generated to reproduce the ramp-up and ramp-down behavior described in \citet{yeamans_high_2021}. The radial laser intensity profile for the radiation-hydrodynamics simulations, $I(r)$, is represented by a super-Gaussian function:
\begin{equation}
    \label{eq:intensity_profile}
    I(r) = I_0 \exp\!\left[-\left(\frac{r}{r_0}\right)^n\right],
\end{equation}
where $I_0$ is the peak intensity, $r$ is the radial distance from the beam center, $r_0 = \SI{500}{\micro\meter}$ is the laser spot size, and $n = 8$ is the super-Gaussian order. The lasers were linearly ramped to full power (\SI{425}{\tera\watt}) over \SI{1.7}{\nano\second} and then deramped to zero power over approximately \SI{200}{\pico\second} after energy delivery to the target; see \cref{fig:laser_profile}.
\noindent For numerical robustness, multiple passes of a binomial smoothing operation were applied to the laser power profile to ensure smooth transitions between the ramp segments.

Because inverse-bremsstrahlung heating followed by electron thermal conduction is the primary laser-to-capsule energy-transfer path, the implosion is highly sensitive to the electron flux limiter $f_{l,e}$. We therefore ran \texttt{xRAGE} for several values of $f_{l,e}$ and tuned it to the measured bang time (\SI{4.36}{\nano\second}). 

Flux limiter values of $f_{l,e} = 0.0275$ and $0.03$ yielded bang times of $\sim$\SI{4.34}{\nano\second} and $\sim$\SI{4.55}{\nano\second}, respectively, with corresponding DT neutron yields of $6.61\times10^{16}$ and $5.90\times10^{16}$. With the bang time matching the measurement to within \SI{20}{\pico\second}, the results obtained using $f_{l,e} = 0.0275$ were adopted to initialize and drive the subsequent \texttt{iFP} simulation.
%
%
\subsection{\texttt{iFP} simulation setup}
\label{subsubsec:ifp_simulation_setup}
\begin{figure}[t]
    \centering
    \includegraphics[scale=0.56]{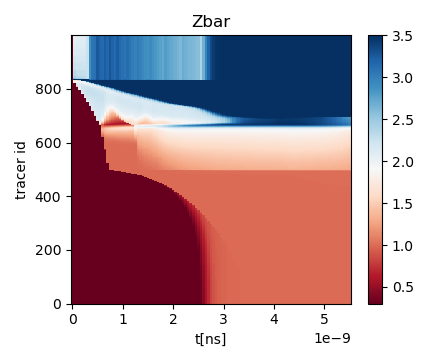}
    \includegraphics[scale=0.56]{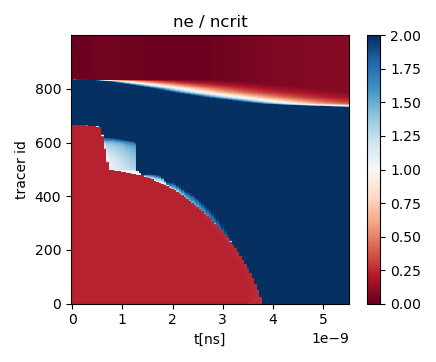}\\
    \includegraphics[scale=0.42]{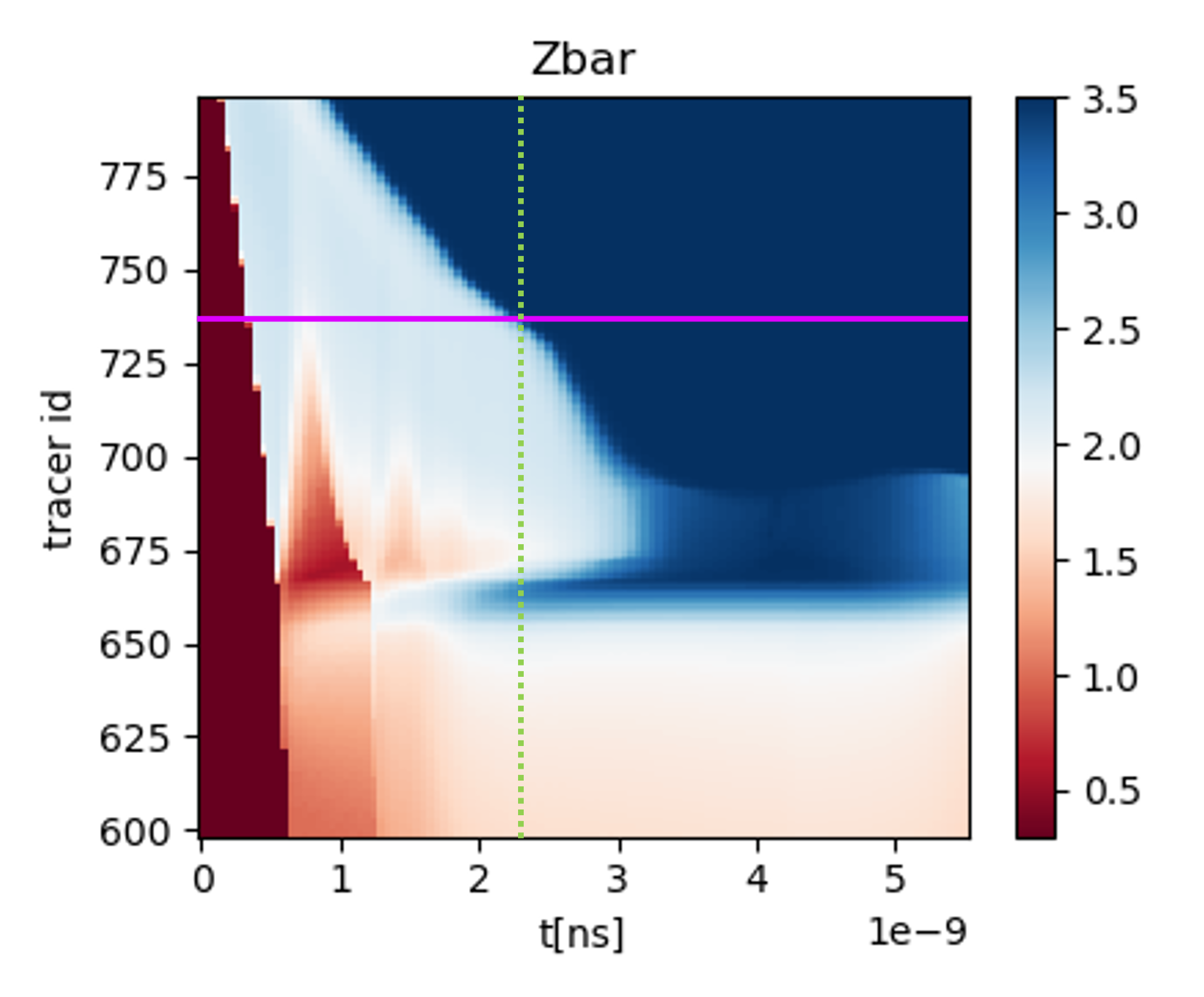}
    \includegraphics[scale=0.42]{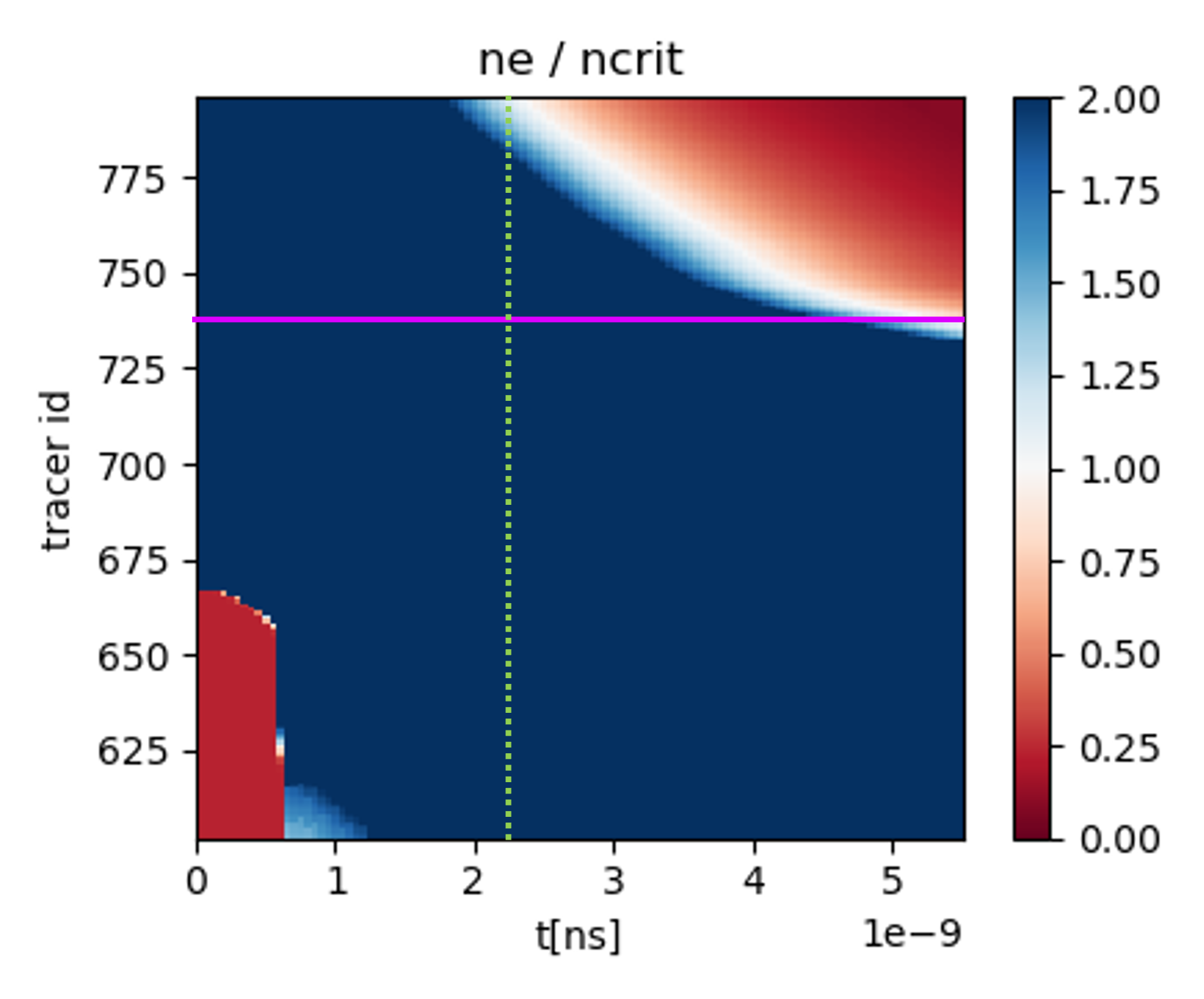}
    \caption{Top: \texttt{xRAGE} results for the average ionization state, $\bar{Z}$ (left), and the ratio of electron number density to the critical density, $n_e/n_{\text{crit}}$ (right), as functions of tracer ID. Bottom: Zoomed-in views of the upper plots between tracer IDs 600 and 800. Tracer ID 739 is highlighted in magenta, and the green vertical line denotes $t_{0,\text{iFP}} = \SI{2.14}{\nano\second}$.}
    \label{fig:xrage_zbar_and_ne_over_ncrit}
\end{figure}

The \texttt{xRAGE} diagnostics are obtained from tracer particles that provide spatiotemporal profiles of various physical quantities. The \texttt{iFP} simulations are initialized using \texttt{xRAGE} output at the time when the fuel adjacent to the ablator has become fully ionized \emph{and} the ionization state within the pusher remains approximately constant, denoted as $t_{0,\text{iFP}}$. The cutoff location of the \texttt{iFP} simulation domain, $r_{\text{iFP}}$, is placed at the quarter-critical-density point on the inner flank of the shell, i.e., radially inside the critical surface and separated from it by overdense ($n_e > n_{\text{crit}}$) shell material. Because the laser cannot propagate past the critical surface, the corona region where laser absorption and associated LPI effects are prominent remains entirely outside the \texttt{iFP} domain and continues to be handled by \texttt{xRAGE}. \Cref{fig:xrage_zbar_and_ne_over_ncrit} shows the \texttt{xRAGE} profiles of the average ionization state and the ratio of electron number density to the critical density.
\noindent At $t_{0,\text{iFP}} = \SI{2.14}{\nano\second}$, a substantial fraction of the fuel zones (ending near tracer ID~500) adjacent to the pusher are fully ionized ($\bar{Z} = 1$). Tracer~739 (hereafter Tr-b) indicates that for $t \geq t_{0,\text{iFP}}$, the ionization state in the surrounding zones remains approximately constant in time. Furthermore, examining the ratio $n_e/n_{\text{crit}}$ shows that Tr-b lies on the inner flank of the shell, shielded from the laser by the overdense layer outside it, ensuring that the laser deposition zone is excluded from the \texttt{iFP} computational domain. Throughout the remainder of the laser pulse ($t_{0,\text{iFP}} \leq t \leq t_{\text{laser}} = \SI{3.9}{\nano\second}$), overdense shell material persists between Tr-b and the critical surface, so the laser deposition zone stays outside the \texttt{iFP} domain while the laser is on. After shutdown, the location of the critical surface is immaterial.

\begin{figure}[t]
    \centering
    \includegraphics[scale=0.6]{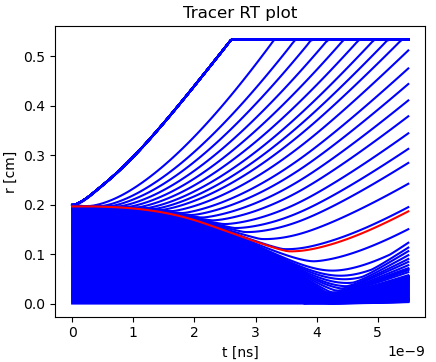}
    \caption{Radius--time (RT) plot of \texttt{xRAGE} tracer particles. The red line indicates the time-dependent cutoff tracer location, $r_{\text{iFP}}(t)$, which defines the boundary of the \texttt{iFP} simulation domain.}
    \label{fig:tracer_rt}
\end{figure}

\begin{figure}[t]
    \centering
    \includegraphics[scale=0.55]{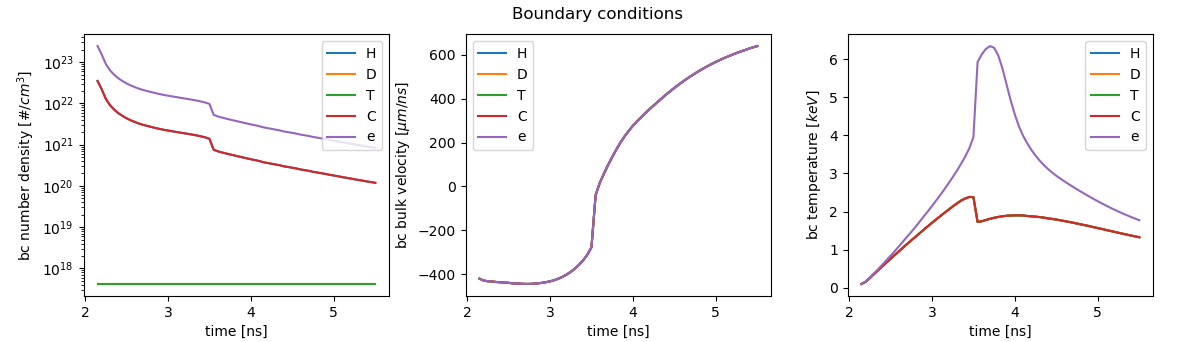}
    \caption{Time-dependent \texttt{iFP} boundary conditions derived from \texttt{xRAGE} Tr-b for (left) number density of each ion species and electrons, (center) radial velocity, and (right) temperature. The number densities of deuterium and tritium are set to negligible values since the boundary cutoff lies within the pusher region. Note that the single-fluid rad-hydro approximation in \texttt{xRAGE} assumes equilibrated ion drift velocities and ion temperatures across all species.}
    \label{fig:ifp_bc}
\end{figure}

A Lagrangian radius--time (RT) streak plot for the \texttt{xRAGE} tracer particles is shown in \cref{fig:tracer_rt}.
\noindent Using the prescribed time-dependent cutoff radius $r_{\text{iFP}}(t)$, we extract the corresponding \texttt{xRAGE} state variables---number density, drift velocity, and ion/electron temperature---to construct a Maxwellian ion distribution function that serves as the boundary condition for the \texttt{iFP} simulation:
\begin{equation}
    \label{eq:ifp_boundary_maxwellian}
    f_\alpha\!\left(r_{\text{iFP}}(t), \vec{v}, t \right)
    = f_M\!\left(\vec{v}; n_{\alpha,b}, \vec{u}_{b}, T_{b}, m_\alpha\right),
\end{equation}
where $f_\alpha(r_{\text{iFP}}, \vec{v}, t)$ is the distribution function of species $\alpha$ at the moving boundary $r_{\text{iFP}}(t)$ (provided by \texttt{xRAGE}), and $n_{\alpha,b}$, $\vec{u}_{b}$, and $T_{b}$ are the time-dependent fluid state variables obtained from Tr-b. The Maxwellian velocity distribution function is given by
\begin{equation}
    \label{eq:maxwellian}
    f_M\!\left(\vec{v}; n, \vec{u}, T, m\right)
    = \frac{n}{\left(2\pi T / m\right)^{3/2}}
      \exp\!\left[-\frac{m}{2T} \left|\vec{v} - \vec{u}\right|^2 \right],
\end{equation}
The resulting time-dependent boundary conditions for the \texttt{iFP} simulation, derived from Tr-b, are shown in \cref{fig:ifp_bc}.
We model the Cutie capsule with \texttt{iFP} over the interval from \SI{2.14}{\nano\second} to approximately \SI{5}{\nano\second}, spanning a period immediately after shock breakout and into the disassembly phase; bang time ($4.36$~ns) occurs while the shock is still diverging outward from the capsule center, before it reaches the ablator--fuel interface and the capsule decelerates, consistent with true exploding-pusher behavior. The time-step size is dynamically adapted according to the heuristics of \citet{taitano_eulerian_2021}, which resolve the dynamical time scales of thermal and momentum equilibration together with the acoustic CFL condition, subject to a maximum step size of $\Delta t_{\text{max}} \sim \SI{0.2}{\pico\second}$. Under this adaptive stepping, with $(\epsilon_{HOLO}, \epsilon_{HO}, \epsilon_{LO}) = (10^{-3},5\times10^{-2},10^{-4})$, the HOLO, HO, and LO iteration counts per step remained bounded at 2--6, 5--7, and 10--20, respectively, throughout the simulation, similar to the Mach-45 case (\cref{fig:m45_verification:solver_stats}). The converged simulations were run on a grid of $(N_x, N_\|, N_\perp) = (384, 128, 64)$, where $N_x$ is the number of radial cells and $N_\|$, $N_\perp$ are the parallel and perpendicular velocity-grid resolutions, respectively.
%
%
%
\subsection{Analysis}
\label{subsubsec:ifp_simulation_analysis}
We compare three simulations: (1) \texttt{xRAGE}, (2) \texttt{iFP} without radiation, and (3) \texttt{iFP} with radiation. \Cref{tab:results_comparison_cutie} summarizes several key observables compared against experimental results.
\begin{table}[t]
    \centering
    \begin{tabular}{ccccc}
        \hline
        \hline
        Model / Experiment & $Y_{DT}$ & $t_{\text{bang}}$ (ns) & $\Delta t_{\text{burn}}$ (ps) & $CR_{\text{vol}}$ \\
        \hline
        \hline
        \texttt{xRAGE} & $6.61\times10^{16}$ & 4.34 & 136 & 1788 \\
        \hline
        \texttt{iFP} without radiation & $4.62\times10^{16}$ & 4.35 & 232 & 395 \\
        \hline
        \texttt{iFP} with radiation & $2.50\times10^{16}$ & 4.41 & 238 & 612 \\
        \hline
        Experiment & $1.19\times10^{16}$ & 4.36 & 421 & 425 \\
        \hline
        \hline
    \end{tabular}
    \caption{Comparison of DT-neutron yield $Y_{DT}$, bang time, burn width, and volumetric convergence ratio $CR_{\text{vol}}$ among the three simulations and the experiment.}
    \label{tab:results_comparison_cutie}
\end{table}
As shown, progressing from \texttt{xRAGE} with ion concentration diffusion to \texttt{iFP} with full ion kinetics and radiation transport systematically reduces the discrepancy with experiment in both the DT-neutron yield and the burn width. The volumetric convergence ratio, $CR_{\text{vol}} = V_{\text{fuel},0} / V_{\text{fuel}}$, also improves dramatically relative to \texttt{xRAGE} ($CR_{\text{vol}} = 1788$, radial $CR_{\text{vol}}^{1/3} = 12.1$): both \texttt{iFP} cases lie close to the experimental value of $425$ (radial $7.5$), with $395$ (radial $7.3$, without radiation) and $612$ (radial $8.5$, with radiation) bracketing it. Note, however, that the $CR_{\text{vol}}$ trend with radiation is non-monotone, so only the yield and burn width improve monotonically with fidelity. The increase in convergence with radiation arises because radiative losses cool the fuel, lowering its pressure and allowing it to converge further. Here, $V_{\text{fuel}}$ denotes the instantaneous volume of the fuel region at stagnation time, defined in more detail shortly.

Although the bang time is slightly delayed in \texttt{iFP} relative to both \texttt{xRAGE} and experiment, this quantity is sensitive to the electron flux limiter in the underlying \texttt{xRAGE} radiation-hydrodynamics simulation that supplies the time-dependent boundary conditions, and could be adjusted accordingly.

\begin{figure}[t]
    \centering
    \includegraphics[scale=0.8]{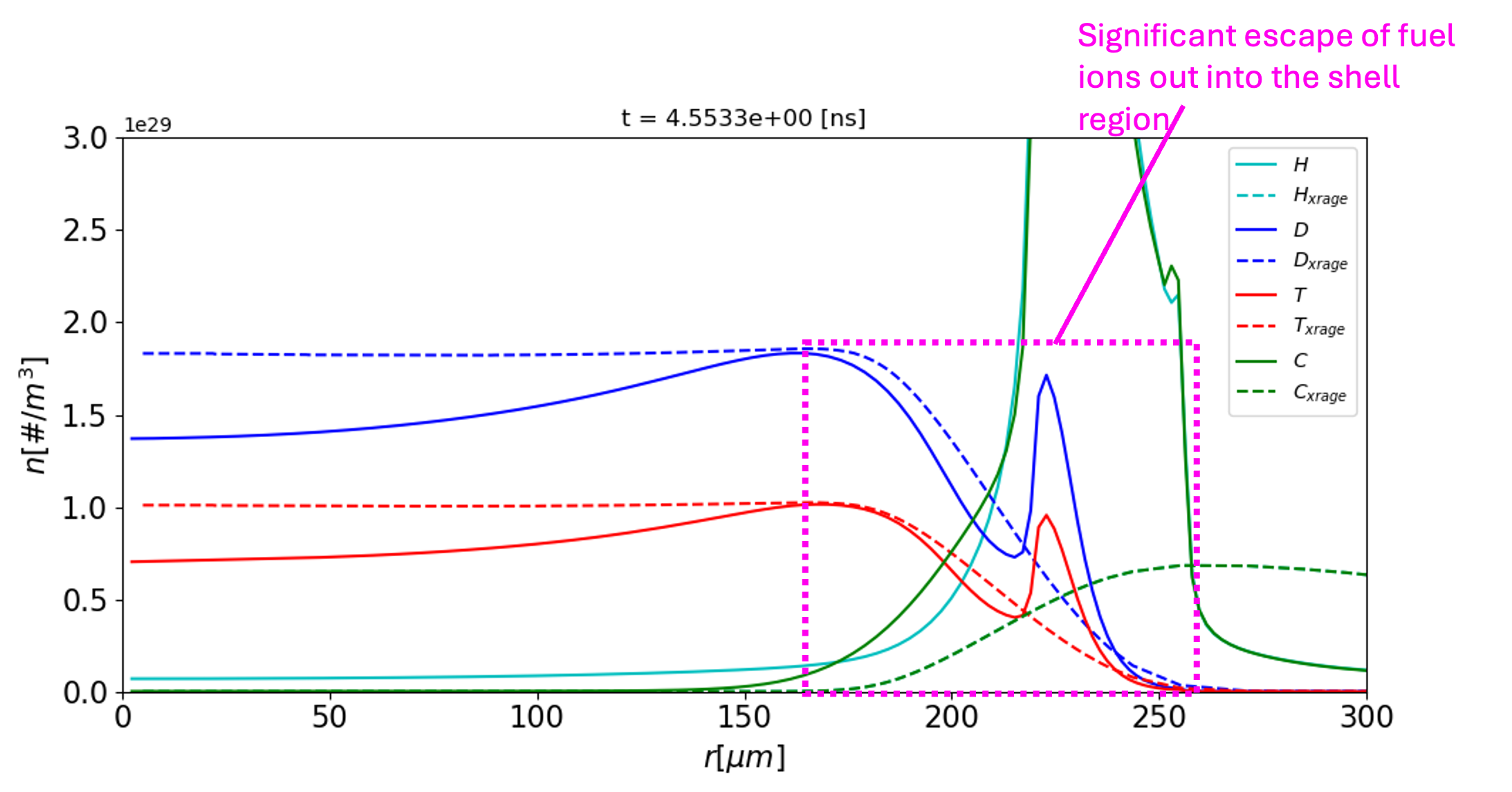}
    \caption{Comparison of ion number densities near bang time between \texttt{iFP} (solid lines) and \texttt{xRAGE} (dashed lines). Species colors as in the legend; hydrogen (cyan) originates from the CH shell.}
    \label{fig:density_depletion_from_core}    
\end{figure}

The trend of improved agreement between higher-fidelity models and experiment can be primarily attributed to two kinetic effects: (1) enhanced fuel-species depletion from the hot spot and (2) increased preheating of the fuel, which raises the ion temperature but reduces convergence---competing effects on the yield.
%
%
%
\paragraph{Kinetically enhanced fuel depletion from the hot spot}
As a consequence of the large fuel Knudsen number (\cref{fig:knudsen_from_xrage}), the fuel particles are expected to deviate significantly from hydrodynamic behavior. \Cref{fig:density_depletion_from_core} compares the ion number densities near bang time between the \texttt{xRAGE} and \texttt{iFP} (with radiation) simulations.
As shown, the \texttt{iFP} density profile exhibits substantially greater interspecies mixing near the fuel--shell interface. The hydrogen density (cyan) penetrates nearly to the center of the domain. To further quantify this, we estimate the characteristic mixing width using a simple diffusion model.

The characteristic diffusive width is approximated as $L_{\text{diff},\alpha} = \sqrt{2\langle D_{\alpha} \rangle \tau}$, where $\tau = t_{\text{bang},\text{iFP}} - t_{0,\text{iFP}} \approx \SI{2.3}{\nano\second}$ represents the effective time window for diffusive mixing. The time-averaged diffusion coefficient is defined as
\begin{equation*}
        \langle D_{\alpha} \rangle =
    \frac{\displaystyle \int_{t_{0,\text{iFP}}}^{t_{\text{bang},\text{iFP}}} \!\! dt
    \left(
        \frac{
            \int_0^{r_b(t)} n_{DT} n_{CH} D_{\alpha}\, dr
        }{
            \int_0^{r_b(t)} n_{DT} n_{CH}\, dr
        }
    \right)}
    {\displaystyle \int_{t_{0,\text{iFP}}}^{t_{\text{bang},\text{iFP}}} \!\! dt}.
\end{equation*}
The instantaneous and local diffusion coefficient for species $\alpha \in \{DT, CH\}$ is estimated as
$D_{\alpha} \sim v_{\text{th},\alpha}^2 \tau_{\text{col},\alpha}$, where
$v_{\text{th},\alpha} = \sqrt{2T_{\alpha}/m_{\alpha}}$ is the thermal speed and
$\tau_{\text{col},\alpha} \propto \left(\sum_{\beta} \nu_{\alpha\beta}\right)^{-1}$ is the effective collision time. The species index $\beta \in \{DT, CH, e\}$ and the collision frequency is given by
$\nu_{\alpha\beta} \propto \frac{Z_{\alpha}^2 Z_{\beta}^2 n_{\beta}}{T_{\beta}^{3/2} m_{\alpha}^{1/2}}$.
The total fuel and shell number densities are $n_{DT} = n_D + n_T$ and $n_{CH} = n_C + n_H$, respectively.

\begin{figure}[t]
    \centering
    \includegraphics[scale=0.6]{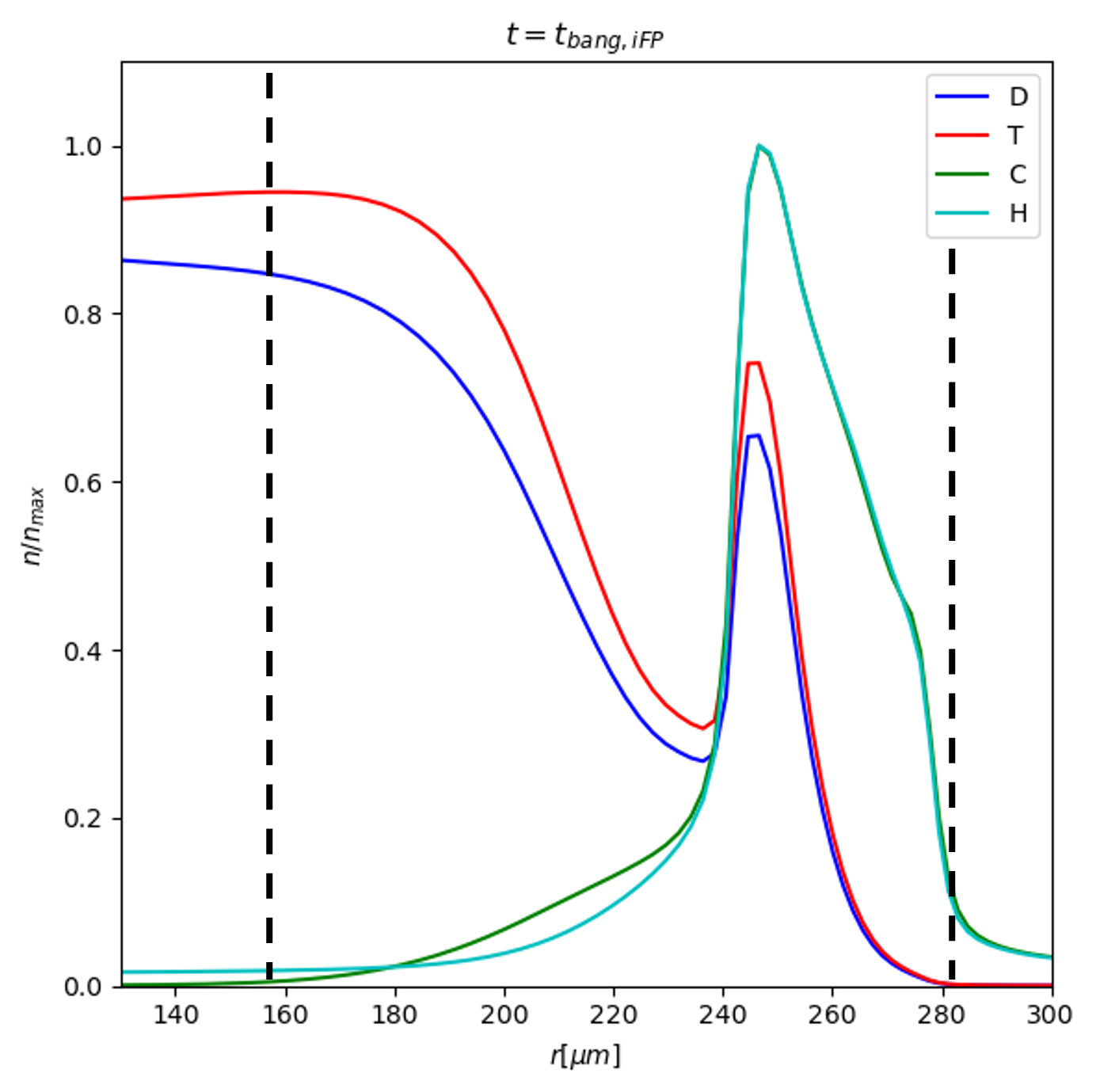}
    \caption{Normalized ion number densities near bang time at the fuel--shell interface from \texttt{iFP}. The interface width is qualitatively indicated by the vertical black dashed lines.}
    \label{fig:density_interface}    
\end{figure}

We find that $L_{\text{diff},DT} = \SI{126}{\micro\meter}$ and $L_{\text{diff},CH} = \SI{21}{\micro\meter}$, giving an average value of $L_{\text{diff,avg}} = \tfrac{1}{2}(L_{\text{diff},DT} + L_{\text{diff},CH}) = \SI{73}{\micro\meter}$.
\Cref{fig:density_interface} shows the normalized ion number densities from \texttt{iFP}.
The observed interface width exceeds the theoretical diffusion estimate. Moreover, the diffusion estimate is itself an upper bound, since it neglects interface compression by the reflected shock. This enhanced interspecies mixing at the interface can be interpreted as a form of \emph{fuel leakage}, where fuel ions that would otherwise participate in fusion reactions escape into the shell region, thereby reducing the overall yield.

To quantify the fuel leakage, we compute the burn-averaged fuel number density defined as
\begin{equation*}
        \langle n_{DT} \rangle_{\text{burn}} =
    \frac{
        \displaystyle \int_{t_{0,\text{iFP}}}^{t_{\text{bang},\text{iFP}}} \!\! dt
        \int_{0}^{r_b(t)} r^2 dr\, R_{DT}\, n_{DT}
    }{
        \displaystyle \int_{t_{0,\text{iFP}}}^{t_{\text{bang},\text{iFP}}} \!\! dt
        \int_{0}^{r_b(t)} r^2 dr\, R_{DT}
    }.
\end{equation*}
Here, $R_{DT}$ denotes the DT reactivity, computed self-consistently by convolving the particle distribution functions with the Bosch--Hale DT reaction cross section~\cite{bosch_hale_1992}. For \texttt{xRAGE} and \texttt{iFP}, we obtain
$\langle n_{DT} \rangle_{\text{burn}} = \SI{7.24e29}{\per\cubic\meter}$ and \SI{1.87e29}{\per\cubic\meter}, respectively,
supporting the qualitative observation of fuel species leakage into the shell region.

Although fuel leakage contributes significantly to yield degradation due to the quadratic dependence of the reaction rate on fuel density, $Y_{DT} \propto n_{DT}^2$ at fixed temperature and burn volume, it is not the full story. Taking the ratio of the squared burn-averaged ion densities between \texttt{xRAGE} and \texttt{iFP} gives a factor of $\sim15$, far exceeding the simulated yield ratio of $\sim2.64$ (\cref{tab:results_comparison_cutie}). Leakage alone would therefore overpredict the yield degradation; the difference is offset by kinetically enhanced preheating, which raises the fuel temperature and burn volume relative to rad-hydro predictions even as it reduces convergence.

\paragraph{Early-time ballistic population and enhanced preheating}

At the time of handoff from \texttt{xRAGE} to \texttt{iFP}, at $t_{0,\text{iFP}}$, the inner surface of the shell possesses an inward velocity of $u_{0,\text{shell}}$ and impacts a fuel region that is approximately at rest. Because the fuel immediately adjacent to the shell is at a much lower density, the interaction between the light fuel ions and the heavy shell can be approximated as scattering off a massive piston. 

A simple kinematic argument shows that a particle with an initial velocity component parallel to the surface normal, $v_{\|}$, scatters off the moving piston with a velocity
$v_{\|,s} = -v_{\|} + 2u_{0,\text{shell}} \approx 2u_{0,\text{shell}}$.
Provided the collisional mean free path of the fuel ions is sufficiently long, and given that the Coulomb relaxation time scale for energetic particles scales as $\sim v_{\|,s}^3$, these scattered particles form a ballistic population capable of streaming significantly ahead of any hydrodynamic bulk motion. Since $v_{\|,s} \approx 2u_{0,\text{shell}}$ is, to leading order, independent of ion mass, the effective kinetic energy of these wall-scattered beams scales as $m_i u_{0,\text{shell}}^2$, so heavier ion species acquire proportionally larger effective temperatures; this is likely the origin of the ion-temperature-versus-species-mass scaling observed experimentally.~\cite{rinderknecht_ion_mass_2015, kabadi_ion_mass_2023} In the extremely kinetic regime considered here, where the interspecies equilibration time scale is long compared to the burn width, a signature of these effective, mass-dependent temperatures is expected to persist into bang time.

\begin{figure}[t]
    \centering
    \includegraphics[scale=0.6]{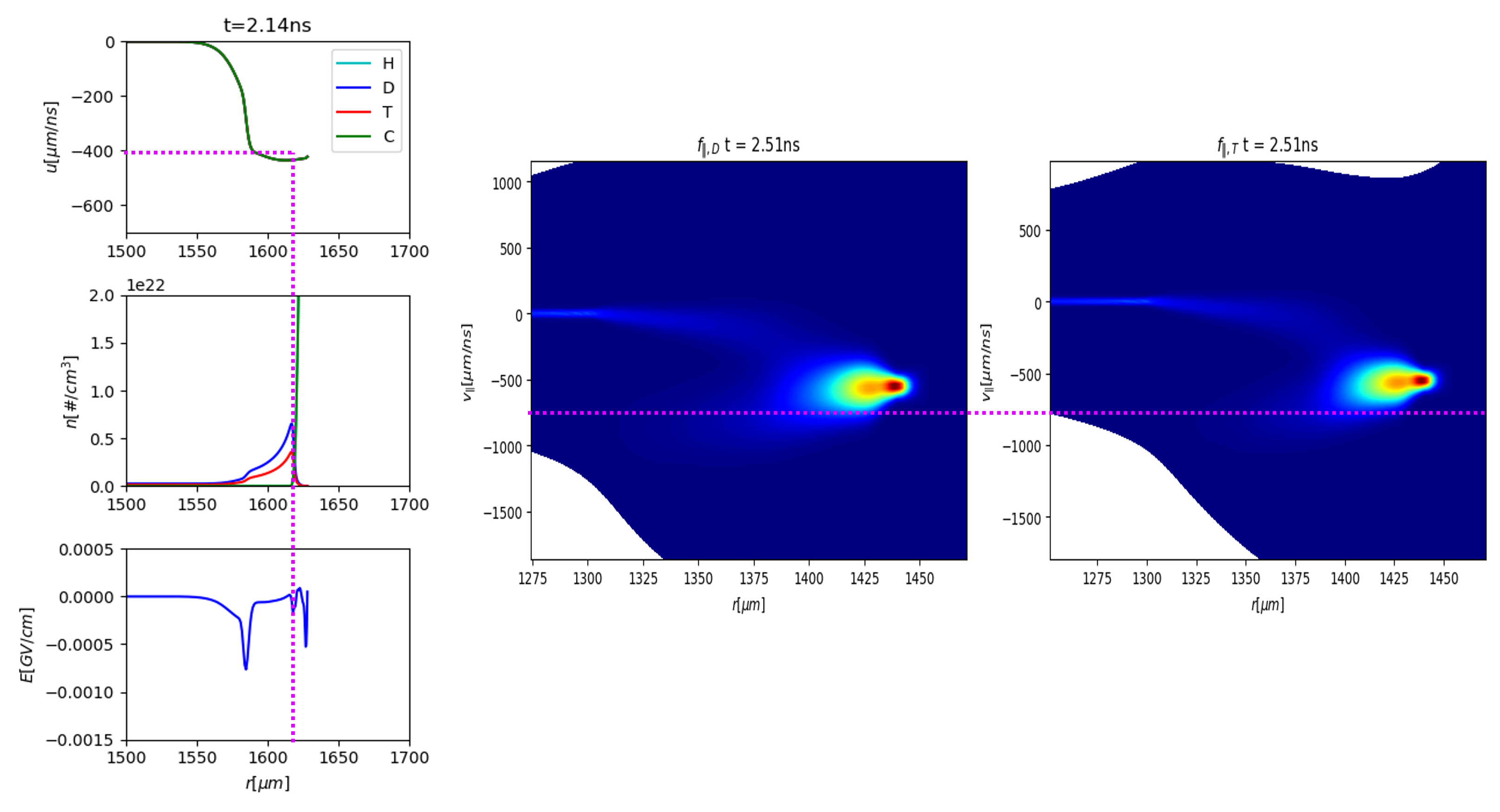}
    \caption{Early-time plasma response near the fuel--pusher interface at \texttt{iFP} initialization ($t \sim \SI{2.14}{\nano\second}$). Shown are (left) the fluid moments---velocity (top), number density (center), and electric field from Ohm's law (bottom)---and (right) the marginal ion distribution functions shortly after initialization ($t \sim \SI{2.5}{\nano\second}$).}
    \label{fig:wall_scattering_demo}
\end{figure}

\Cref{fig:wall_scattering_demo} shows the early-time structure of the fuel ion (D, T) marginal particle distribution functions (PDFs), along with their corresponding velocity and density moments, and the associated electric field.
As seen in the right panel, a bimodal structure develops in the marginal PDF, with the high-speed beam component (highlighted by magenta dashed lines) centered around $\sim 2u_{0,\text{shell}}$. For $u_{0,\text{shell}} \approx \SI{-400}{\micro\meter\per\nano\second}$, this behavior confirms the physical expectation for reflected particle velocities. Minor deviations from this idealized prediction arise from electrostatic acceleration and from the finite shell mass, which allows partial momentum exchange.

\begin{figure}[t]
    \centering
    \includegraphics[scale=0.8]{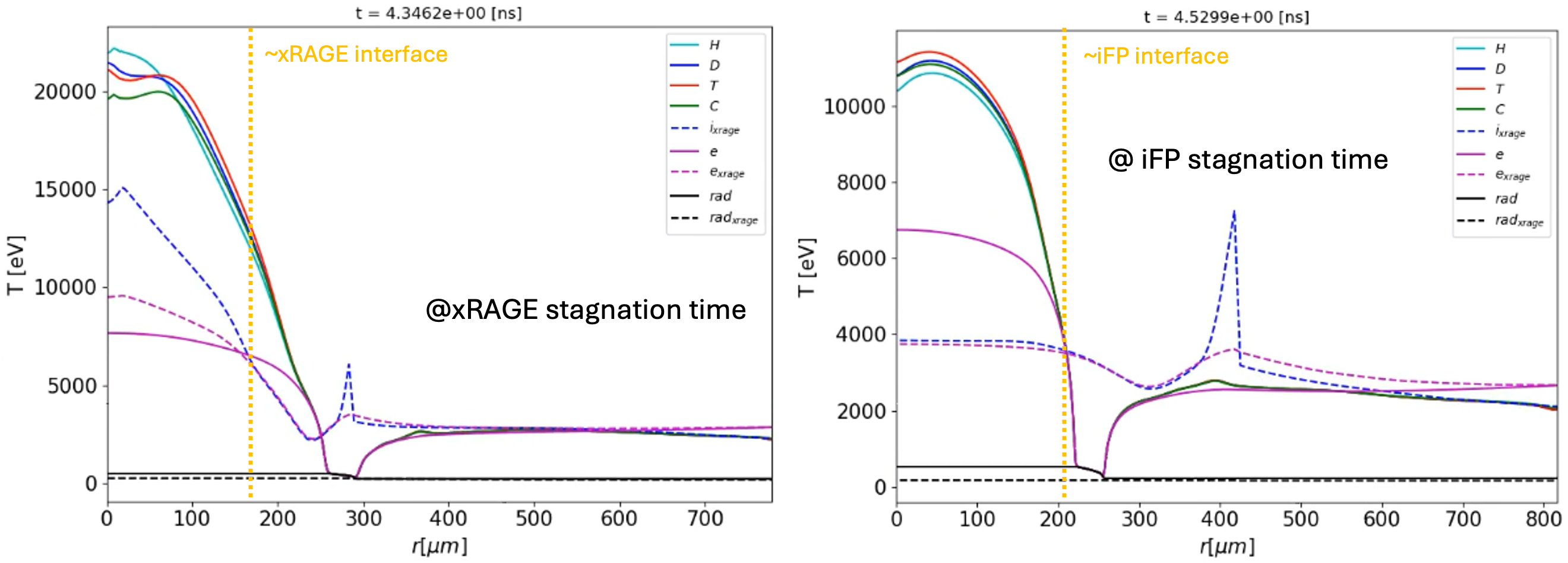}
    \caption{Temperature profiles near bang time for \texttt{xRAGE} (left) and \texttt{iFP} (right). The larger stagnation radius in \texttt{iFP} reflects the effects of kinetic preheating.}
    \label{fig:xrage_ifp_interface_temperature_bangtime}
\end{figure}

\begin{figure}[t]
    \centering
    \includegraphics[scale=0.7]{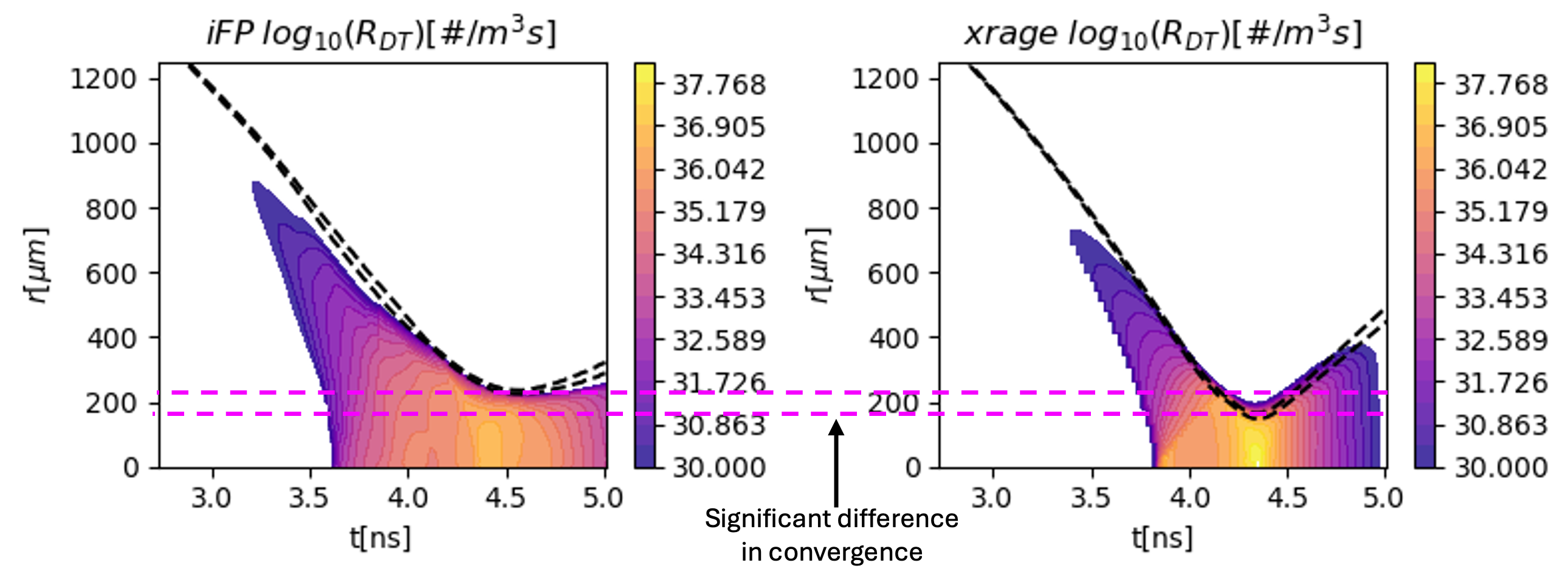}
    \caption{RT streak plots of the DT reactivity for \texttt{iFP} (left) and \texttt{xRAGE} (right). The black dashed lines indicate the 90\% (inner) and 98\% (outer) fuel containment radii. The magenta lines denote the minimum 90\% containment positions for each simulation.}
    \label{fig:reactivity_rt_comparison}
\end{figure}

The early-time ballistic ion population propagates deep into the core, leading to preheating and an increase in the fuel adiabat. This effect is illustrated in \cref{fig:xrage_ifp_interface_temperature_bangtime}, which compares the bang-time temperature and interface locations between \texttt{xRAGE} and \texttt{iFP}. Both codes allow for interspecies mixing: \texttt{xRAGE} through an ion concentration diffusion model, whose validity is restricted to small Knudsen number, and \texttt{iFP} through the self-consistent, kinetic evolution of ion species PDFs, which remains valid regardless of Knudsen number. In this study, we define the fuel--shell interface location as the radius that encloses 95\% of the total fuel content, determined by
$N_{\text{fuel}}(r_{95\%}) = 4\pi \int_{0}^{r_{95\%}} r^2 \sum_{i \in \{D, T\}} n_i\, dr$ (the 90\% and 98\% contours in the streak plots bracket this interface).
As seen, energetic-ion-induced preheating causes a larger stagnation radius in \texttt{iFP} relative to \texttt{xRAGE}. The dynamic impact of this behavior is further illustrated in \cref{fig:reactivity_rt_comparison}, which compares radius--time streak plots of the DT reactivity between the two simulations.
The \texttt{iFP} simulation exhibits a more gradual stagnation process, a larger minimum radius, and an extended burn duration, consistent with the experimentally observed increase in burn width. The reactivity-weighted fuel temperatures, time-averaged over the burn width, are
$\langle T_{DT} \rangle_{\text{burn}} = \SI{11.50}{\kilo\electronvolt}$ for \texttt{xRAGE} and \SI{12.13}{\kilo\electronvolt} for \texttt{iFP}, confirming that ballistic heating mechanisms enhance the fuel temperature in the kinetic model.

Finally, we note that radiation energy partitioning and transport are essential for quantitative agreement. When radiation is excluded (\texttt{iFP} without radiation), the plasma retains more of the deposited energy: the fuel is hotter and converges less ($CR_{\text{vol}} = 395$ vs.\ 612), and the yield is overpredicted by nearly a factor of two relative to the radiative case. This follows from the strong ($\propto T^4$) scaling of the radiation energy density: when radiation is present, an increasing share of the deposited energy is partitioned into the radiation field as the plasma heats, thereby limiting the plasma temperature rise relative to the case without radiation.

In summary, progressively increasing physics fidelity---from \texttt{xRAGE} with ion-concentration diffusion, to \texttt{iFP} with ion kinetics alone, to \texttt{iFP} with ion kinetics \emph{and} radiation---systematically improves agreement with the experimentally measured yield and burn width. Radiation transport plays an essential, not incidental, role: without it, the fuel retains too much energy, and the yield is overpredicted.

%% file: sections/conclusions.tex
\section{Conclusions and Future Directions}
\label{sec:conclusions}

We have presented, to our knowledge, the first coupling of a kinetic-ion plasma
model to thermal radiation transport. The \texttt{iFP} code---a conservative,
spherically symmetric, 1D2V hybrid kinetic-ion/fluid-electron
Vlasov--Fokker--Planck solver---was extended with a grey $P_1$ radiation model,
coupled to the plasma through a high-order/low-order (HOLO) fixed-point
solver~\cite{chacon_multiscale_2017} and advanced on a shared adaptive moving
mesh. Material opacities are assembled on the fly from the LANL \texttt{TOPS} library
using a mass-weighted mixing model that exploits the per-species information
available in a kinetic code. We verified the implementation against the optically thick
Marshak-wave and Mach-45 radiative-shock benchmarks, confirming correct behavior
in both the diffusive, strongly collisional limit and the strong-shock,
material-motion-dominated regime.

Applying the coupled capability to the NIF polar-direct-drive ``Cutie''
exploding-pusher capsule, we found that increasing the physics fidelity---from
\texttt{xRAGE} with ion-concentration diffusion, to \texttt{iFP} with full ion
kinetics, to \texttt{iFP} with ion kinetics \emph{and} radiation---systematically
reduces the discrepancy with experiment in both the DT-neutron yield and the
burn width. We attribute this improvement to two ion-kinetic mechanisms that a
hydrodynamic description misses: kinetically enhanced fuel-species leakage out of
the hot spot, and preheating of the fuel by an early-time ballistic ion
population, which raises the fuel temperature and adiabat but reduces
convergence. Importantly,
radiation transport is essential for quantitative agreement: with radiation
omitted, the fuel retains too much energy, converges less, and the yield is
overpredicted by nearly a factor of two. To our knowledge, this is the first demonstration of an ICF implosion
modeled with a self-consistently coupled kinetic-ion and radiation description,
and it establishes the coupling machinery on which higher-fidelity treatments can
be built.

These results come with clear caveats that frame the next steps. The radiation
field was restricted to a single (grey) energy group. This choice was deliberate:
a grey $P_1$ closure is the simplest model that still exercises the full
coupling---absorption/emission exchange, material-motion corrections, and the
shared moving mesh---and it lets us isolate the role of radiation in the
implosion without conflating it with spectral effects. It is nonetheless a
quantitative limitation, because the radiation spectrum in these capsules is expected
to depart strongly from a Planckian; a multigroup treatment is required for spectral fidelity.
A second limitation is a modeling inconsistency at the boundary: the
time-dependent drive is supplied by an \texttt{xRAGE} simulation that uses grey
radiation \emph{diffusion}, whereas \texttt{iFP} evolves the grey $P_1$ moments,
so the two codes do not treat the radiation field identically across the handoff.
Reconciling the driver and interior radiation models is a priority for
quantitative predictions. Finally, several physics packages remain absent from
\texttt{iFP} for fully self-consistent, end-to-end capsule modeling---most
notably a laser package, non-ideal and strongly coupled ion equation-of-state
models, and multigroup radiation transport---which is why the present study
relies on an \texttt{xRAGE}-driven boundary condition rather than modeling the
laser--plasma interaction directly.

Immediate next steps are to extend the grey $P_1$ model to a multigroup
treatment and to implement more sophisticated opacity models. In the longer term
we intend to replace the moment-based radiation closure with a fully kinetic
radiation-transport treatment, drawing on our in-house kinetic-radiation
solvers~\cite{park_multigroup_2019, hammer_multi-dimensional_2019}, and to couple
the kinetic-electron capability already present in
\texttt{iFP}~\cite{anderson_efficient_2020} to the radiation field. Together with
a laser package and improved equation-of-state models, these developments define
a path toward a first-principles, end-to-end kinetic simulation
capability for ICF and high-energy-density experiments.

%% file: sections/appendix.tex
\section{Moving-mesh transformation and discretization of the radiation moments}
\label{app:moving_mesh}

This appendix collects the details deferred from
\cref{subsec:model:moving_mesh,subsec:model:moving_mesh_equations,sec:solver}:
the coordinate transformation that puts the governing equations in
arbitrary-Lagrangian--Eulerian (ALE) conservative form on the moving mesh
(\cref{app:mm:transform}), the nonlinear elimination of the $P_1$ radiation flux
(\cref{app:mm:elimination}), and the semi-discrete finite-volume discretization
of the radiation moments (\cref{app:mm:discretization}). The plasma
(ion VFP and electron) equations are transformed and discretized by the same
construction; those details, together with the monitor-function limiting and the
volume-preserving mesh stabilization, are unchanged from
\citet{taitano_cpc_2021} and are not reproduced here.

\subsection{Coordinate transformation}
\label{app:mm:transform}

The mesh is defined by the time-dependent map $r(\xi,t)$ from the fixed logical
coordinate $\xi\in[0,1]$ to the physical radius, with boundaries
$r|_{\xi=0}=0$ and $r|_{\xi=1}=R(t)$ [\cref{eq:model:mesh_map}]. Two
Jacobians appear: the mesh Jacobian $J_\xi = \partial_\xi r$ and the spherical
Jacobian $J_S = r^2$. Derivatives at fixed physical position and at fixed logical
position are related through the grid velocity $\dot{r}\equiv\partial_t r|_\xi$;
for a scalar $\phi$,
\begin{equation}
    \label{eq:app:time_transform}
    \partial_t \phi\big|_{r} = \partial_t \phi\big|_\xi
    - \frac{\dot{r}}{J_\xi}\,\partial_\xi \phi,
    \qquad
    \partial_r \phi = \frac{1}{J_\xi}\,\partial_\xi \phi .
\end{equation}
The advection term $-(\dot{r}/J_\xi)\,\partial_\xi\phi$ accounts for the motion of
the grid relative to fixed physical positions. Applying \cref{eq:app:time_transform} to the
grey $P_1$ energy equation \cref{eq:model:radiation:energy}, written in physical
coordinates as
$\partial_t(J_S E_r) + \partial_r(J_S F_\|) + J_S S_{re} = 0$, and using the
geometric conservation law
\begin{equation}
    \label{eq:app:gcl}
    \partial_t J_\xi\big|_\xi = \partial_\xi \dot{r},
\end{equation}
the product rule yields the conservative ALE form
\begin{equation}
    \label{eq:app:mm_energy}
    \partial_t\!\left(J_\xi J_S\, E_r\right)
    + \partial_\xi\!\left[J_S\!\left(F_\| - \dot{r}\,E_r\right)\right]
    + J_\xi J_S\, S_{re} = 0,
\end{equation}
which is \cref{eq:model:mm:radiation_energy}. The flux (momentum) equation
\cref{eq:model:radiation:momentum} carries the additional factor $1/c^2$ on the
time derivative and the isotropic radiation-pressure gradient $\third\partial_r
E_r$. Transforming the time derivative as in \cref{eq:app:time_transform} and
the gradient as $\partial_r E_r = J_\xi^{-1}\partial_\xi E_r$, multiplying the
equation through by $J_\xi$, and using \cref{eq:app:gcl} to combine
$-E_r\,\partial_\xi\dot{r}-\dot{r}\,\partial_\xi E_r = -\partial_\xi(\dot{r}E_r)$, gives
\begin{equation}
    \label{eq:app:mm_flux}
    \frac{1}{c^2}\partial_t\!\left(J_\xi J_S\, F_\|\right)
    - \frac{1}{c^2}\partial_\xi\!\left(J_S\,\dot{r}\,F_\|\right)
    + \third J_S\,\partial_\xi E_r
    + J_\xi J_S\, S_{rp,\|} = 0,
\end{equation}
i.e.\ \cref{eq:model:mm:radiation_flux}. Because the radiation pressure is
isotropic, the pressure term enters as the gradient $\third J_S\,\partial_\xi E_r$
rather than a conservative divergence; the geometric source terms that would
accompany a divergence cancel. The ion VFP equation \cref{eq:model:ion:vfp} and
the electron energy equation \cref{eq:model:electron:energy} transform in exactly
the same way, producing \cref{eq:model:mm:vfp,eq:model:mm:electron_energy}.

\subsection{Nonlinear elimination of the $P_1$ flux}
\label{app:mm:elimination}

The flux equation \cref{eq:app:mm_flux} can be solved for $F_\|$ and eliminated
from the system, reducing the radiation unknowns to the energy $E_r$ alone.
Writing the BDF2 time derivative of the flux as
$\partial_t\!\left(J_\xi J_S F_\|\right) = \left(a_0 J_\xi J_S F_\|
+ \sum_{\ell\ge1} a_\ell J_\xi^{\,l+1-\ell} J_S^{\,l+1-\ell} F_\|^{\,l+1-\ell}\right)/\Delta t^{l}$,
where $a_\ell$ are the BDF2 weights (in the notation of
\cref{eq:solver:bdf2_stencil}, $a_0 \equiv a^{\,l+1}$ and
$a_\ell \equiv a^{\,l+1-\ell}$), and inserting the comoving relation
\cref{eq:model:radiation:comoving} and the deposition closure
\cref{eq:model:radiation:simplified}, the momentum
balance gives
\begin{equation}
    \label{eq:app:p1_flux}
    J_\xi J_S\left(\dfrac{\sigma_r}{c} + \dfrac{a_0}{c^2\,\Delta t^{l}}\right) F_\|
    - \frac{1}{c^2}\partial_\xi\!\left(J_S\,\dot{r}\,F_\|\right)
    =
    -\sum_{\ell\ge1}\dfrac{a_\ell}{c^2\,\Delta t^{l}}\,J_\xi^{l+1-\ell} J_S^{l+1-\ell} F_\|^{\,l+1-\ell}
    - \third\,J_S\,\partial_\xi E_r
    + J_\xi J_S\dfrac{4}{3}\dfrac{\sigma_r}{c}\,u_{e,\|} E_r.
\end{equation}
Because of the mesh-advection term, \cref{eq:app:p1_flux} couples neighboring
faces and forms a tridiagonal system for $F_\|$, which is solved before the
flux is substituted into the energy equation.
Eliminating $F_\|$ requires a corresponding modification of the Jacobian when the
system is solved by a Newton method. Retaining only the radiation block with
unknowns $(E_r, F_\|)$, the Jacobian has the structure
\begin{equation}
    \label{eq:app:jacobian_block}
    J =
    \begin{bmatrix}
        J_{EE} & J_{EF} \\
        J_{FE} & J_{FF}
    \end{bmatrix},
\end{equation}
and eliminating the flux unknowns yields the Schur complement
$\tilde{J}_{EE} = J_{EE} - J_{EF}J_{FF}^{-1}J_{FE}$.
In practice \texttt{iFP} uses the linearized update
\begin{equation}
    \label{eq:app:linearized_flux}
    J_\xi J_S\left(\dfrac{\sigma_r}{c} + \dfrac{a_0}{c^2\,\Delta t^{l}}\right)\delta F_\|
    - \frac{1}{c^2}\partial_\xi\!\left(J_S\,\dot{r}\,\delta F_\|\right)
    =
    - \third\,J_S\,\partial_\xi\!\left(\delta E_r\right)
    + J_\xi J_S\tfrac{4}{3}\dfrac{\sigma_r}{c}\!\left(\delta u_{e,\|} E_r + u_{e,\|}\,\delta E_r\right),
\end{equation}
where $\delta(\,\cdot\,)$ denotes the Newton increment; $\delta F_\|$ is then
substituted into the linearized energy equation.

\subsection{Semi-discrete finite-volume discretization}
\label{app:mm:discretization}

The moving-mesh equations are discretized with a conservative finite-volume
scheme on the logical grid: cell index $i$, faces $i\pm\half$, logical cell width
$\Delta\xi_i$, and time-level superscript $l$ (omitted when the current level is
implied). The mesh Jacobian $J_\xi$ and spherical Jacobian $J_S$ are evaluated
per cell/face, with
\begin{equation}
    \label{eq:app:JS_cases}
    J_S =
    \begin{cases}
        1, & \text{planar},\\
        r^2, & \text{spherical},
    \end{cases}
\end{equation}
so the same code path covers planar and spherical geometry. Face values are
obtained from cell values by a projection operator $\upwind{\,\cdot\,}_{i+\half}$;
for a first-order upwind choice,
\begin{equation}
    \label{eq:app:upwind}
    \upwind{q, u}_{i+\half} =
    \begin{cases}
        q_i,     & u_{i+\half}\ge 0,\\
        q_{i+1}, & u_{i+\half}<0,
    \end{cases}
\end{equation}
with higher-order limited variants (SMART~\cite{gaskell_curvature-compensated_1988}, WENO5~\cite{jiang_efficient_1996}) used in practice. The discrete grey $P_1$
energy equation \cref{eq:app:mm_energy} is
\begin{multline}
    \label{eq:app:discrete_energy}
    \frac{1}{\Delta t^{l}}\sum_{\ell} a_\ell\, J_{\xi,i}^{\,l+1-\ell} J_{S,i}^{\,l+1-\ell} E_{r,i}^{\,l+1-\ell}
    - \frac{J_{S,i+\half}\,\dot{r}_{i+\half}\upwind{E_r,\dot{r}}_{i+\half}
    - J_{S,i-\half}\,\dot{r}_{i-\half}\upwind{E_r,\dot{r}}_{i-\half}}{\Delta\xi_i} \\
    + \frac{J_{S,i+\half}\,F_{\|,i+\half} - J_{S,i-\half}\,F_{\|,i-\half}}{\Delta\xi_i}
    + J_{\xi,i} J_{S,i}\, S_{re,i} = 0,
\end{multline}
and the discrete flux equation \cref{eq:app:mm_flux} is given below. The
cell-centered fluxes in the mesh-motion term are reconstructed as
$F_{\|,i} = \half\left(F_{\|,i-\half} + F_{\|,i+\half}\right)$, which couples
$F_{\|,i\pm\half}$ to $F_{\|,i\pm\frac{3}{2}}$ and yields the tridiagonal system
noted in \cref{app:mm:elimination}:
\begin{multline}
    \label{eq:app:discrete_flux}
    J_{\xi,i+\half} \left(\frac{a_0}{c^2\,\Delta t^{l}} + \frac{\sigma_{r,i+\half}}{c}\right) F_{\|,i+\half}
    - 
    \frac{1}{c^2}\,\frac{\dot{r}_{i+1} F_{\|,i+1} - \dot{r}_i F_{\|,i}}{\Delta\xi_{i+\half}} \\
    = -\frac{1}{c^2\,\Delta t^{l}}\sum_{\ell\ge1} a_\ell\, J_{\xi,i+\half}^{\,l+1-\ell} F_{\|,i+\half}^{\,l+1-\ell}
    - \third\,\frac{E_{r,i+1} - E_{r,i}}{\Delta\xi_{i+\half}}
    + \tfrac{4}{3} J_{\xi,i+\half} \frac{\sigma_{r,i+\half}}{c}\,u_{e,i+\half}\upwind{E_r,u_e}_{i+\half}.
\end{multline}
In \texttt{iFP} we adopt the non-conservative form of the flux equation, which removes the coupling to $J_S$ and improves robustness; when absorption and re-emission are significant, we observe the resulting non-conservation error to be small empirically. Cross sections are harmonically averaged between neighboring cells. The energy deposition and the momentum coupling (through the Ohm's-law electric field, \cref{eq:model:lo_electrical_field}) are discretized as
\begin{align}
    \label{eq:app:discrete_deposition}
    S_{re,i} &= \sigma_{a,i}\, c\, E_{r,i} - \sigma_{e,i}\, a c\, T_{e,i}^4
    - u_{e,i}\,\third\,\frac{E_{r,i+1} - E_{r,i-1}}{\Delta r_{i+\half} + \Delta r_{i-\half}}, \\
    \label{eq:app:discrete_momentum}
    S_{rp,\|,i+\half} &= -\third\,\frac{E_{r,i+1} - E_{r,i}}{\Delta r_{i+\half}},
    \qquad
    S_{rp,\|,i} = \tfrac{1}{2}\!\left(S_{rp,\|,i+\half} + S_{rp,\|,i-\half}\right).
\end{align}
We use Marshak boundary conditions for the flux on the boundary faces,
\begin{equation}
    \label{eq:app:marshak_bc}
    F_{\half} = \frac{c}{2}\left(E_0 - E_1\right) + \frac{4}{3}u_{e,\half}\upwind{E_r,u_e}_{\half},
    \qquad 
    F_{N+\half} = \frac{c}{2}\left(E_N - E_{N+1}\right) + \frac{4}{3}u_{e,N+\half}\upwind{E_r,u_e}_{N+\half},
\end{equation}
where $E_0$ and $E_{N+1}$ are the prescribed left and right boundary (ghost)
radiation energy densities.